\documentclass{jfm}

\usepackage{graphicx}
\usepackage{natbib}
\usepackage{amssymb}
\usepackage{upmath}
\usepackage{amsbsy}
\usepackage{mdwlist}
\usepackage{fancyhdr}

  \usepackage[breaklinks=true, colorlinks=true, citecolor=blue, linkcolor=blue, urlcolor=blue, bookmarksdepth=3, pdftitle={The importance of bubble deformability for strong drag reduction in bubbly turbulent Taylor--Couette flow}]{hyperref}

\fancypagestyle{firststyle}
{

\fancyhf{}{Published in \emph{J. Fluid Mech.} (2013), \emph{vol.} 722, \emph{pp.} 317--347.\\ doi: \href{http://dx.doi.org/10.1017/jfm.2013.96}{10.1017/jfm.2013.96}\vspace{-20mm}}
\fancyfoot{}
}

\newcommand\x{\times}
\newcommand\gvf{\alpha_{global}}
\newcommand\amax{\alpha_{max}}
\newcommand\ramax{r_{\alpha_{max}}}
\newcommand\We{We}
\newcommand\Fr{Fr_{cent}}

\title[The importance of bubble deformability for strong DR in turbulent TC flow]{The importance of bubble deformability for strong drag reduction in bubbly turbulent Taylor--Couette flow}

\author[D. P. M. van Gils, D. Narezo Guzman, C. Sun and D. Lohse]%
{Dennis P.\ M.\ van Gils\footnote[2]{Present address: Max Planck Institute for Dynamics and Self-Organization, 37077
G\"ottingen, Germany.}, Daniela Narezo Guzman, Chao Sun\footnote[1]{Email addresses for correspondence: \href{mailto://c.sun@utwente.nl}{c.sun@utwente.nl}, \href{mailto://d.lohse@utwente.nl}{d.lohse@utwente.nl}}\linebreak and Detlef Lohse$\dagger$}

\affiliation{Physics of Fluids Group, Faculty of Science and Technology, J.M. Burgers Center for Fluid Dynamics, and IMPACT Institute, University of Twente, The Netherlands}

\pubyear{2013}
\volume{722}
\pagerange{317--347}
\date{28 November 2011; revised 1 October 2012; accepted 12 February 2013;\linebreak first published online 28 March 2013}

\begin{document}
\maketitle
\thispagestyle{firststyle}


\begin{abstract}

\noindent
Bubbly turbulent Taylor--Couette (TC) flow is globally and locally studied at Reynolds numbers of $Re = 5 \x 10^5$ to $2 \x 10^6$ with a stationary outer cylinder and a mean bubble diameter around 1 mm. We measure the drag reduction (DR) based on the global dimensional torque as a function of the global gas volume fraction $\gvf$ over the range 0--4$\%$. We observe a moderate DR of up to $7\%$ for $Re = 5.1 \x 10^5$. Significantly stronger DR is achieved for $Re = 1.0 \x 10^6$ and $2.0 \x 10^6$ with, remarkably, more than $40\%$ of DR at $Re = 2.0 \x 10^6$ and $\gvf = 4\%$.

To shed light on the two apparently different regimes of moderate DR and strong DR, we investigate the local liquid flow velocity and the local bubble statistics, in particular the  radial gas concentration profiles and the bubble size distribution, for the two different cases: $Re = 5.1 \x 10^5$ in the moderate DR regime and $Re = 1.0 \x 10^6$ in the strong DR regime, both at $\gvf = 3 \pm 0.5\%$ .

In both cases the bubbles mostly accumulate close to the inner cylinder (IC). Surprisingly, the maximum local gas concentration near the IC for $Re =1.0 \x 10^6$ is $\approx 2.3$ times lower than that for $Re =5.1 \x 10^5$, in spite of the stronger DR. Evidently, a higher local gas concentration near the inner wall does not guarantee a larger DR.

By defining and measuring a local bubble Weber number ($\We$) in the TC gap close to the IC wall, we observe that the cross-over from the moderate to the strong DR regime occurs roughly at the cross-over of $\We \sim 1$. In the strong DR regime at $Re = 1.0 \x 10^6$ we find $\We > 1$, reaching a value of 9 (+7, -2) when approaching the inner wall, indicating that the bubbles increasingly deform as they draw near the inner wall. In the moderate DR regime at $Re = 5.1 \x 10^5$ we find $\We \approx$ 1, indicating more rigid bubbles, even though the mean bubble diameter is larger, namely 1.2 (+0.7, -0.1) mm, as compared with the $Re = 1.0 \x 10^6$ case, where it is
0.9 (+0.6, -0.1) mm. We conclude that bubble deformability is a relevant mechanism behind the observed strong DR. These local results match and extend the conclusions from the global flow experiments as found by \citet{ber05} and from the numerical simulations by \citet*{lu05}.


\bigskip\hrule\bigskip

\end{abstract}
\pagebreak

\section{Introduction} \label{sec:introduction}

Theoretical, numerical and experimental studies on drag reduction (DR) of a solid body moving in a turbulent flow have been performed for more than three decades. DR can be achieved by using, e.g., surfactants \citep{sae00, dra06}, polymers \citep{vir75,ber78,benzi04,bonn05,whi08,pro08} and bubbles \citep{mad84,mad85,cla91,Lat97,kat98,tak00,ber05,murai05,deu06,san06,ber07,Elb08,gut08,mur08,sug08b,jacob10,cec10}.
The proper implementation and understanding of DR through bubbles is specially relevant for naval applications, since it can lead to significant reduction of the fuel consumed by ships without adding substances into water. Despite all efforts spent in understanding the fundamental mechanisms behind this effect, a solid understanding of the DR mechanisms occurring in bubbly flows is still missing. Several mechanisms have been proposed as relevant, for example, bubble splitting \citep{meng98}, bubble compressibility \citep{lo06}, bubble deformability \citep{ber05, lu05} and effective compressibility of the flow \citep{fer04}. The different mechanisms as listed here do not necessarily exclude each other and are likely to be active simultaneously. Depending on the Reynolds number $Re$, to be defined in \S\ref{sec:setup}, the bubble size, the bubble deformability and other parameters, one mechanism could dominate over the other(s) \citep{mer92,mor02,she06}. A frequently occurring categorization on DR studies as found in the literature, is based on the bubble size. Studies on micrometre- and (sub)millimetric-sized bubbles, commonly injected in or near the boundary layer of interest, have received a lot of attention in the last few decades (see, e.g., \cite{mad84,mad85,mer89,cla91,mer92,kat98,meng98,kod00,kan01,mor02,fer04,murai05,san06,she06,gut08} amongst many already cited references). The experimental results as presented in the current work uses millimeter-sized bubbles injected in the free stream.

The conventional systems for studying bubbly DR are channel flows \citep{kod00}, flat plates \citep{mer89, lat03,san06}, and cavity flows \citep[see the review article by][]{cec10}. A general explanation of the physical mechanism behind DR is complicated since the flow in each type of the aforementioned systems can exhibit either spatial dependence, as in developing boundary layers, or temporal dependence, as in air film build-up and bubble shedding. This and other factors make it difficult to collapse all bubbly DR data from different set-ups, see again the review by \citet{cec10} on experimental work on bubbly DR measured on plates and on the boundary of channel flows. More recently, experiments on Taylor--Couette (TC) flow have been conducted, overcoming this problem since statistically steady states are easy to achieve \citep{djeridi04,ber05, murai05,ber07,meh07,sug08b}. The TC system has the advantage of being a closed system and thus having a well-defined energy balance (see e.g.\ \citet{lat92a} or \citet{eck07b})
in analogy with the Rayleigh-B\'enard system \citep{ahl09}. In addition, TC flow offers the possibility of measuring global torque and gas volume fraction with good accuracy. Taking these various aspects into account, TC flow is ideal for studying the drag that a surface experiences when moving in a bubbly flow.

Bubbly TC flow has been experimentally studied \citep{murai05, meh07} and theoretically modeled \citep{sug08b} for low $Re$ numbers ($< 10^4$). The mechanism responsible for the DR in that regime is that the injected bubbles disrupt the coherent flow structures, leading to DR. The efficiency of this particular mechanism gets lost for high $Re$ numbers when typical fluid velocity fluctuations are larger and coherent structures are increasingly unsteady.

In the high-$Re$-number regime, the experimental work by \citet{ber05} on TC flow employing bubbles on one hand and buoyant non-deformable particles on the other, has proven the existence of two different DR regimes within $7 \x 10^4< Re <10^6$. This work focused only on the flow's global response under influence of bubble/particle injection, treating the changes to the local flow parameters as a black box. The division between the two DR regimes roughly occurs at a Weber number $\We \sim 1$. The Weber number compares the fluid inertia around the bubble surface to the surface tension. In the regime for $Re \lesssim 6 \x 10^5$ and $\We < 1$, the bubbles remain close to spherical and the DR is moderate, found to be around 5\%. The mechanism of effective compressibility of the mixture, suggested in the numerical work by \citet{fer04}, explains the DR obtained in this regime. In the regime for larger $Re$ we have $\We > 1$ and the DR is significant, reaching up to around 20\%. In this regime effective compressibility becomes less dominant and the relevant mechanism is related to the deformability of the bubbles (which are much larger than the Kolmogorov length scale $\eta_K$). In support of this view, \citet{ber05} find that using non-deformable buoyant particles of the same size leads to little DR. This interpretation has also been suggested in the numerical work by \citet*{lu05}. From their simulations these authors conclude that the deformable bubbles close to the wall push high-vorticity regions far away from the wall, reducing the drag. They suggest that this pushing is caused by the lift force on the bubble, which strongly depends on its shape \citep{mag00}.

In a later work on the same TC setup, \citet{ber07} compared the bubbly DR for smooth versus rough walls within $1 \x 10^5 < Re < 4 \x 10^5$, again focussing on the flow's global response. Their results support the claim that bubbly DR is a boundary layer effect as bubbly DR totally vanishes for the rough-wall case.

Significant bubbly DR with smooth walls has been measured by \citet{ber05} in a narrow $Re$ regime from $6 \x 10^5$ to $1 \x 10^6$. Given above interpretation on bubble deformability, it is expected that a system capable of larger $Re$ numbers should lead to more DR. This assumption needs to be validated. Furthermore, previous DR measurements were based on the global torque, leaving many issues of the mechanism responsible for this effect unknown: How do bubbles modify the liquid flow? How do they distribute themselves inside the gap? What is the local Weber number of the bubbles? To gain more insight, \emph{local} information on the flow and on the bubbles is required.

In the present work we explore the $Re$ parameter space with inner cylinder rotation up to $Re = 2.0 \x 10^6$ and we measure the DR as function of the global gas volume fraction $\gvf$. For $Re = 5.1 \x 10^5$ and $1.0 \x 10^6$ we investigate the local modifications to the azimuthal velocity of the fluid due to the bubbles, we measure the local distribution of the bubbles and their sizes, and we present their local Weber number. These two $Re$ numbers are chosen because the smaller one corresponds to the moderate DR regime and the larger one to the strong DR regime.

The outline of the paper is as follows. The experimental setup and global measurement techniques are introduced in \S \ref{sec:setup}, the local measurement techniques and their accuracy in \S \ref{sec:local_techniques}. The results on the global parameters are presented in \S \ref{sec:global_exp} and the local parameters in \S \ref{sec:local_exp}. Finally, conclusions are drawn in \S \ref{sec:conclusion}, followed by further discussions and an outlook in \S \ref{sec:discussion}.


\begin{figure}
\center
\includegraphics[height=7cm]{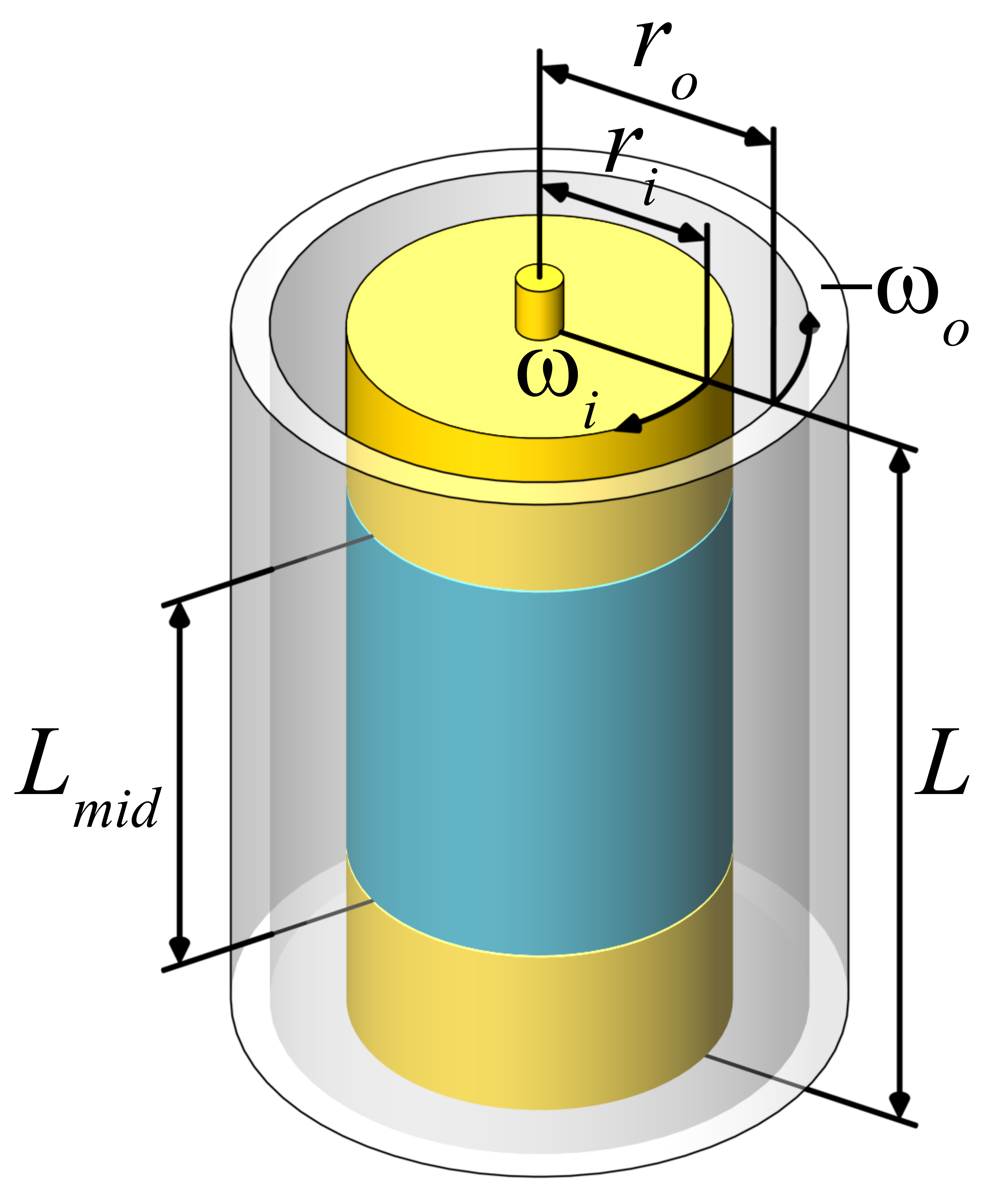}
\caption{Schematic view of TC flow and its important parameters. In this work we only focus on the torque acting on the middle section of the inner cylinder (IC), shown in blue. The outer cylinder (OC) is kept at rest.
\label{fig:01}}
\end{figure}

\begin{table}
\centering
\begin{tabular}[c]{c c c c c c c}
$r_i$  & $r_o$  & $L$   & $L_{mid}$ & $\Gamma$ & $\eta$ & $V_{gap}$\\
$(m)$  & (m)    & (m)   & (m)       &          &        & (m$^3$)\\
0.2000 & 0.2794 & 0.927 & 0.536     & 11.68    & 0.716  & 0.111\\
\\
\end{tabular}
\caption{Geometric parameters of the present TC system; the T$^3$C. Here, $r_i$ and $r_o$ are the radii of the inner and outer cylinders, respectively, $L$ is the length of the full IC, $L_{mid}$ is the length of the middle section of the IC and $V_{gap}$ is the total volume of the TC gap. The aspect ratio is given by $\Gamma\equiv L/(r_o - r_i)$ and the radius ratio by $\eta\equiv r_i/r_o$.
\label{table:T3C_dimensions}}
\end{table}

\section{Experimental setup and global measurement techniques} \label{sec:setup}


\subsection{Experimental setup}

The experiments are performed in the Twente turbulent Taylor--Couette (T$^3$C) facility, see \citet{gil11a} for a detailed description of the system. Here we only give a brief introduction on the system parameters. We refer to figure \ref{fig:01} for a schematic view of the set-up and to table \ref{table:T3C_dimensions} for the values of the geometric parameters. The inner cylinder (IC) of the T$^3$C is divided into three separate sections. The height of the middle section of the IC is $L_{mid} = 0.536$ m, being $58\%$ of the full cylinder height $L=0.927$ m. The IC has a radius of $r_i = 0.2000$ m and the outer cylinder (OC) a radius of $r_o = 0.2794$ m. The maximal inner and outer angular velocities are $\omega_i/2\pi = 20$ Hz and $\omega_o/2\pi = \pm 10$ Hz, respectively. We use the radial distance $r$, that transforms into the non-dimensional gap distance $\tilde{r}=(r - r_i)/(r_o - r_i)$ with 0 indicating the IC wall and 1 the OC wall. The system is fully temperature controlled through cooling of the upper and lower plates. In the present work we only rotate the IC and we use decalcified water as the working fluid and filtered instrument air for the bubble injection.

The Reynolds number $Re$ for pure IC rotation flow is given by
\begin{equation}
Re=U_i(r_o-r_i)/\nu
\end{equation}
\noindent where $U_i=\omega_i r_i$ is the azimuthal velocity of the IC wall, $\omega_i$ is its angular velocity and $\nu$ is the kinematic viscosity of the liquid.


\subsection{Torque measurement}

As shown in figure \ref{fig:01}, the IC of the T$^3$C is divided into three separate sections (top, middle and bottom), each designed to measure the torque by virtue of a load cell utilizing strain-gauge deformation. Each section is basically a hollow cylinder suspended onto the IC's drive axle by two low-friction ball bearings. A metal arm connects the drive axle to the inner wall of each IC section. Each arm has a split in the middle that is bridged by a load cell \citep[see figure 7 in][]{gil11a}. The torque measurements are performed on the middle section of the IC, that were shown to have minimal end plate effects for the turbulent case of pure IC rotation \citep{gil12}.


\subsection{Bubble injection and global gas volume fraction measurement}

Eight upright bubble injectors are equally distributed around the perimeter of the bottom plate. Each injector consists of a capillary, housed inside a plug ending flush with the interior wall. In all of our measurements we employed the same capillaries of 1 mm in diameter. The gas injection rate is regulated by two mass flow controllers, resulting in 216 liters per minute at the maximum injection rate. The injected bubble size depends only on the shear stress in the flow for the $Re$ range examined here \citep{ris98}.

\begin{figure}
\center
\includegraphics[height=8cm]{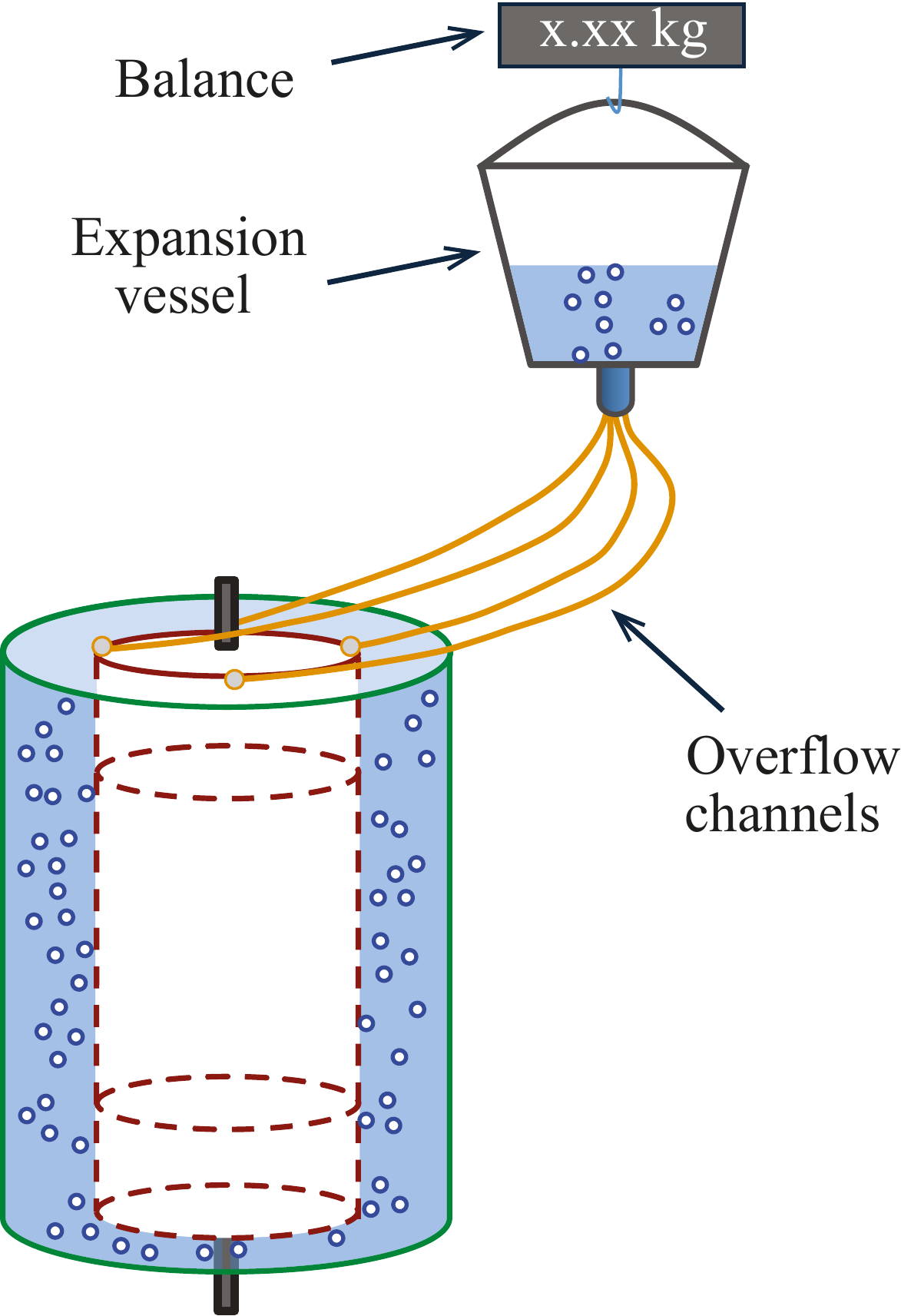}
\caption{Schematic view of the set-up showing the overflow channels through which excess water and air exit. The ejected water is collected in the bucket, that is continuously weighed by the balance, providing $\gvf$.
\label{fig:02}}
\end{figure}

When air injection is activated, the excess of liquid and air needs to go out of the tank to avoid pressure build up. Water and air escape through four exits, referred to as overflow channels, connecting the inner volume of the TC, denoted by $V_{gap}$, to the outside. The overflow channels are located on the top plate, close to the IC. Hoses are attached to them and lead to a bucket, where the escaping water is collected. A balance registers the weight of the suspended bucket over time at a sampling rate of 5 Hz, see figure \ref{fig:02} (see also figure \ref{fig:05} later).

We set the weight of the bucket for single phase flow as the weight zero offset. Once air injection is activated and the two-phase flow achieves stability over time, i.e. the bucket's weight is constant over time, the air volume contained inside $V_{gap}$ can be calculated. Considering the continuously acquired temperature of the flow, the density of the water at each moment is known and thus the gas volume in the tank. The global gas volume fraction $\gvf$ is given by the gas volume in the gap as percentage of $V_{gap}$.

The top, middle and bottom IC sections have small gaps in between them to decouple the torque acquisition of the middle section to the neighbouring sections. These small gaps introduce an increasing error over time in the global gas volume fraction estimation. The centripetal force pushes the bubbles through the separations and into the cavities between the sections. Part of the water that initially was inside, is pushed eventually into the flow and contributes to the fluid collected in the bucket, thus overestimating the actual global gas volume fraction. We conduct experiments to quantify a maximal error of 0.5\% for a global gas concentration of 3\%, and the error is smaller for smaller gas concentrations.


\section{Local measurement techniques} \label{sec:local_techniques}


\subsection{Azimuthal liquid velocity measurement - LDA} \label{sec:ch7_LDA_technique}

Laser Doppler Anemometry (LDA) is used to study the fluid's azimuthal velocity component of the single and two-phase flows. We make use of a LDA system from Dantec Dynamics, consisting of a back-scatter probe head (85 mm probe 60X81) with a beam expander (55X21) of 1.98 expansion and an achromatic front lens of 500 mm focal distance. This results in a measurement volume of dimensions width x height x depth = 0.07 mm x 0.07 mm x 0.3 mm. The probe head is mounted on a custom-built computer controlled traversing system, rigidly attached to the T$^3$C frame. The LDA system further consists of two photomultipliers (57X18) and a burst spectrum analyzer (BSA F80). The seeding particles we use are round and have a density of 1.03 g cm$^{-3}$ and a mean diameter of 5 $\mu$m (polyamide seeding particles, PSP-5, Dantec Dynamics). As shown in \cite{gil12}, a force balance estimation tells us that centrifugal forces are at least four orders of magnitude smaller than the drag forces acting on the particles, and hence centrifugal effects are negligible on the particles in our examined flows.

When applying LDA to our set-up, the two laser beams in the horizontal plane refract in a non-linear way due to the curved surface of the OC. Therefore, the angle between them, $\theta$, depends on the radial position of the focal point when located inside the gap between the cylinders. To correct for the curvature effect, not taken into account in the Dantec software, the measured speed is multiplied by the factor
\begin{equation}
C_{\theta}=\frac {n_\mathrm{air}\mathrm{sin}(\theta_\mathrm{air}/2)} {n_\mathrm{water}\mathrm{sin}(\theta_\mathrm{water}/2)}
\end{equation}
where $n_\mathrm{air}$ and $n_\mathrm{water}$ are the refractive indices of air and water, and $\theta_\mathrm{air}$ and $\theta_\mathrm{water}$ are the angles between the beams when found in air or water. We refer to the work of \citet{hui12b} for a more detailed explanation of the LDA method used in our TC set-up.


\subsection{Bubble statistics measurement - Optical fibre probe} \label{sec:fibre_probe_technique}
\subsubsection{Introduction}
To retrieve local information on the bubble statistics inside the flow, such as the local gas concentration and horizontal bubble chord lengths, we employ the optical fibre probe technique. This technique has been used widely in bubbly flows \citep{mud01,xue03,gue03,lut04b,wu08,xue08,mar10,berg11}. The optical probe is a local phase detection device, i.e. able to discriminate between water and air in our case. Light that is coupled into one end of the optical fibre, will be internally reflected back at the other extreme; the ``tip''. The reflection coefficient depends primarily on the refractive index of the phase in which the fibre tip is submerged. By acquiring the amount of reflected light over time, one can obtain the bubble statistics. For a detailed description of the working principle, we refer to \citet{car90,car91}. We also note that optical probes in the flow obviously disturb the flow characteristics and statistics. For the high Reynolds numbers and the small probes here we consider this problem to be minor. For the different flow geometry of the Twente water channel (a vertically recirculating closed channel flow), but at similar Reynolds numbers, \cite{ren05} did not find any effect of the optical probes on the spectra. Still, the issue of the probe back reaction on the flow requires attention for closed systems, such as TC and Rayleigh--B\'enard flows.

\begin{figure}
\center
\includegraphics[width=1\textwidth]{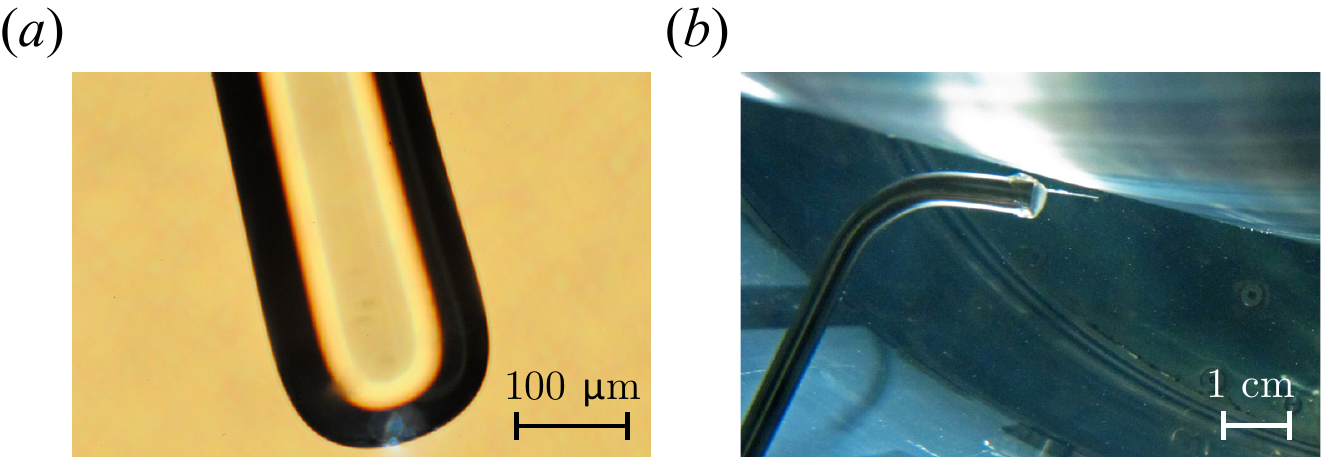}
\caption{(\emph{a}) Microscopic photograph of the fibre tip. (\emph{b}) T$^3$C gap top view: mounted optical fibre probe to measure the bubble statistics close to the IC wall.
\label{fig:03}}
\end{figure}

\begin{figure}
\center
\includegraphics[width=.8\textwidth]{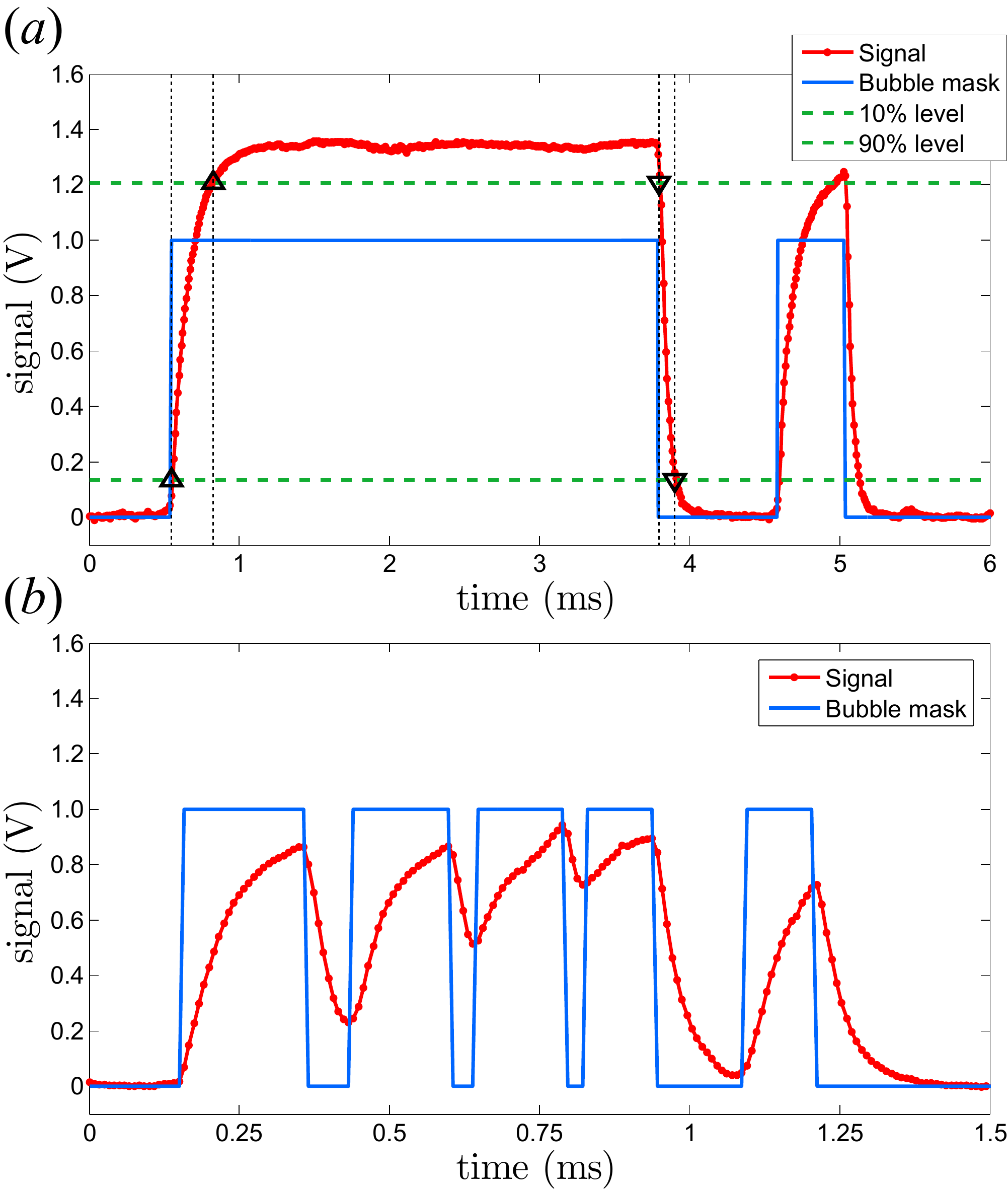}
\caption{Typical detrended voltage signals (red lines with dots) of bubble--probe interactions as acquired by our fibre probe placed inside the T$^3$C facility, (\emph{a}) operated at $Re = 5.1\times 10^5$ and $\gvf = 3\%$, displaying two distinct bubbles, (\emph{b}) operated at $Re = 1.0\times 10^6$ and $\gvf = 3\%$, displaying five distinct bubbles. A positive signal slope indicates dewetting and a negative slope indicates rewetting of the tip. By applying a threshold algorithm on the detrended signal, based on the signal amplitude and its time derivative, a bubble mask is obtained (solid blue lines). The width of the bubble mask is taken as the residence time $T_{bubble}$ per specific bubble. The dashed vertical and horizontal lines and open triangles in panel (\emph{a}) aid the discussion on the latency length $L$ in \S \ref{sec:fibre_probe_technique}. Note that the plotted time range is smaller in panel (\emph{b}). The signal is acquired at a sampling rate of 120 kHz.
\label{fig:04}}
\end{figure}

For our measurements we use a custom-made glass fibre probe (0.37 NA Hard Polymer Clad Multimode fibre, Thorlabs Inc., diameter of $200 \mu$m), whose tip is stripped off the surrounding cladding and is heat treated to have a smooth U-shaped dome, see figure \ref{fig:03}(\emph{a}). The internally reflected light is collected by a photodiode and is electronically amplified (four-point probe electronics, Kramer's Laboratorium, Delft University of Technology). In all our experiments an acquisition card (National Instruments 6221) samples the fibre signal voltage with 16-bit resolution at a sampling rate of 120 kHz. The choice for this specific sampling rate was determined by precursory tests and ensures that the bubble--probe interactions are sufficiently resolved (see figure \ref{fig:04}). The optical probe sticks through a sensor hole of the OC with the fibre tip placed inside the TC gap, parallel to the azimuthal axis. It is mounted on a manual traversing system with a total radial traveling distance of $27 \pm 0.02$ mm along the gap, which is around one third of the gap between the cylinders. Figure \ref{fig:03}(\emph{b}) shows a top view of the gap and the optical probe, made possible due to transparent windows on the end plates of the T$^3$C.

The accuracy of this technique has been studied in water--air flows by \citet{julia05} with bubbles of diameter 2.8--5.2 mm and velocities of 0.22--0.28 m s$^{-1}$, the latter one order of magnitude smaller than in our experiments. The residence time is the time the fibre tip is immersed in one bubble and we refer to it as $T_{bubble}$. It depends on the bubble size, the bubble velocity, and also on the frontal location where the probe tip pierces the bubble. \citet{julia05} report a small underestimation of $T_{bubble}$, that, according to these authors, is caused by the local deformation of the bubble due to probe-induced liquid pressure over the bubble or due to being hit by the probe in itself. The bubble can possibly also be decelerated or deviated from its path by an upstream high-pressure region induced by the probe's presence. In our experiments the bubble velocities are one order of magnitude larger than in \citet{julia05}, and hence they have a larger added mass. While this supports the notion that it is less probable that a bubble is decelerated or deviated from its trajectory by the probe's presence in our flows, proofing this claim is hard. A force balance estimation on these effects remains elusive, due to the complex nature of the coupled bubble--flow interactions. Also, fully spatially and temporally resolved flow fields around the bubbles in the $Re$ regime examined here, are hard, if not impossible with the current state of experimental techniques. Hence, we cannot guarantee that the bubbles will not be deformed, deflected or decelerated due to the presence of the probe, and we will take this into account by amply increasing the error bars of our local bubble statistics measurements, discussed in \S \ref{sec:1pp_accuracy}. As will turn out, even with these error bars we can make qualitative conclusions on the bubble--flow dynamics.

\cite{car90} introduces the concept of a latency length $L$ for a phase-detecting optical fibre probe, resembling the spatial resolution of the interface detection by the given probe. In order to appreciate the concept, we first take a look at an example of a bubble--probe interaction signal as acquired with our probe. Figure \ref{fig:04}(\emph{a}) shows two bubble--probe events taken with our T$^{3}$C facility, operated at $Re = 5.1\times 10^5$ and $\gvf = 3\%$. The signal is detrended such that 0 volts corresponds to the fibre tip submerged in water. We focus our attention on the first bubble--probe interaction, starting around time $t$ = 0.5 ms: a clear plateau is reached in the voltage signal, indicating that the probe has pierced the bubble and has had sufficient residence time in the gas phase for the tip to dewet completely. The time it takes for the signal to go from 10$\%$ to 90$\%$ of the mean voltage making up the plateau, is an experimental determination of the duration of the signal rise $T_m$, as put forward by \cite{car90}. The 10$\%$ and 90$\%$ slice levels are indicated by the dashed horizontal green lines in figure \ref{fig:04}(\emph{a}), with the upwards pointing triangles and corresponding dotted vertical lines signalling the starting and ending time markers of this specific $T_m$. In a similar fashion one can arrive at an estimate for the signal fall time $T_d$ -- commonly smaller than $T_m$ \citep{car90} -- indicated by the solid downwards pointing triangles. By taking $V_i$ as the velocity of the air-water interface hitting the probe, viz.\ the bubble velocity, \cite{car90} arrives at $L=T_m V_i$ as the latency length of the probe. In the next section \S \ref{sec:1pp_validation} we will provide an estimate of $L$ for our current probe.

The bottom line is that $L$ (or likewise, $T_m$ and $T_d$) should be short enough when compared to the size (residence time) of the bubbles, in order to easily resolve the bubble--probe interactions. If the response time $T_m$ becomes too large with respect to the bubble--probe interaction time, either due to too small bubbles or due to too high bubble impact velocities, the fibre tip will not have enough time to dewet completely. This then leads to signals as similar to the second bubble--probe interaction of figure \ref{fig:04}(\emph{a}) and the full figure \ref{fig:04}(\emph{b}), in which we show bubble--probe signals of our most challenging locally investigated bubbly flow at $Re = 1.0\times 10^6$ and $\gvf = 3\%$ at a radial location near the IC, $\tilde{r} = 0.008$. We emphasize that such signals are still usable if appropriate signal post-processing settings are chosen. To evaluate the optical probe against another independent measurement technique, we will compare the estimated horizontal bubble diameter as obtained from the optical probe with  image analysis results of high-speed image recordings, presented in \S \ref{sec:bubble_size}. Nevertheless, an evaluation on the latency length, the signal post-processing settings and the resulting accuracy is in place.

\subsubsection{Validation and post-processing}
\label{sec:1pp_validation}
We obtain the latency length $L$ of our optical fibre probe by evaluating the acquired bubble signals, as will be presented in the upcoming sections. Only bubble--probe interaction signals that exhibit a clear plateau as shown in figure \ref{fig:04}(\emph{a}, first interaction) are selected for the evaluation. After detrending the signal, we calculate the 10$\%$ and 90$\%$ slice levels of each of these events in order to retrieve the average signal rise time $\langle T_m \rangle$ with standard deviation $\sigma_{m}$, and the average signal fall time $\langle T_d \rangle$ with standard deviation $\sigma_{d}$, taken over $N$ such events at a radial distance of $\tilde{r} \sim 0.1$. Finally, we look ahead at the time-averaged measured azimuthal liquid velocity of the two two-phase flow cases at that radial location (see figure \ref{fig:08}), and we set $V_i$ equal to this velocity by assuming no slip or drift of the bubbles. This assumption will be examined in future work. The results are summarized in table \ref{table:1pp_characteristics}. For the case of $Re = 1.0 \times 10^6$ the bubble velocities are larger, leading to less occurrences of complete dewetting of the tip and leading to smaller $T_m$ values. The signal fall time $T_d$, however, is unaffected by the higher bubble impact velocities for the cases examined here. We can only speculate on the cause for this. The resulting latency lengths $L$ are different for the two $Re$ cases and are of order 1 mm, similar  to figure 12 of \cite{car90},  which shows the same order of magnitude and trend for the U-shaped probe with larger latency lengths at higher $V_i$, although the interface velocities examined there are two to five times smaller than in our case.

\begin{table}
\centering
\begin{tabular}[c]{l c c c c c c c}
& $N$ & $\langle T_m \rangle$ & $\sigma_m$ & $\langle T_d \rangle$ & $\sigma_d$ & $V_i$ & $L$\\
& & (ms) & (ms) & (ms) & (ms) & (m s$^{-1}$) & (mm)\\
\\
$Re = 5.1\times10^5$, $\gvf=3\pm0.5\%$: & 685 & 0.31 & 0.09 & 0.10 & 0.01 & 2.9 & 0.9\\
$Re = 1.0\times10^6$, $\gvf=3\pm0.5\%$: & 354 & 0.25 & 0.08 & 0.10 & 0.01 & 5.7 & 1.4\\
\\
\end{tabular}
\caption{Bubble detection characteristics of our optical probe for the two two-phase flow cases as studied locally in the current work. A selection of $N$ bubble--probe interactions that lead to a complete dewetting of the tip is used to estimate the average signal rise time $\langle T_m \rangle$ with standard deviation $\sigma_m$ and the average signal fall time $\langle T_d \rangle$ with standard deviation $\sigma_d$. The bubble--water interface velocity is $V_i$ and $L=T_m V_i$ is the latency length of the optical probe \citep{car90}.
\label{table:1pp_characteristics}}
\end{table}

Evidently, the latency length of our probe is of the same order as our bubble size. This causes many wedge-shaped bubble--probe interaction signals with non-constant local signal maxima and minima, as in figure \ref{fig:04}(\emph{b}). In this figure the first four bubbles are in such short succession after each other, that the tip does not rewet completely and, consequently, that the signal does not drop to zero before a new bubble is being pierced, resulting in a `jagged' signal. Clearly, simple signal amplitude threshold schemes, as discussed in detail by \cite{car90}, are not able to resolve the signal into separate bubble events. The post-processing algorithm we apply, makes use of the first time derivative of the signal to discriminate between tip dewetting and tip rewetting. The complete algorithm is as follows: first, a single signal amplitude threshold is applied on the detrended signal in order to detect a first signal rise above the noise level. This point in time is set as the first bubble arrival time $T_{AT}$. The noise level in all our runs was $\pm 0.04$ V and the applied signal amplitude threshold was $V_\mathrm{thres}$ = 0.08 V, set as low as possible while still safely away from the noise level. Second, we calculate the time-derivative of the signal and find the first point in time $T_1$ after $T_{AT}$ at which the signal rate drops below $-8.4 \times 10^3$ V s$^{-1}$. This threshold value is at 75\% of the signal rate associated with the air-film drainage as obtained from the $T_d$ study. The 75\% is still large enough to neglect signal oscillations that sometimes happen in the case of `overshoot' when reaching a signal plateau, as
explained by \cite{car90}. We also calculate the first point in time $T_2$ after $T_{AT}$ at which the signal amplitude drops below $V_\mathrm{thres}$ again. The bubble exit time is now set to $T_{ET} = \mathrm{min}(T_1, T_2)$. Third, we set the next bubble arrival time to that point in time at which the signal has a positive slope combined with a signal amplitude larger than $V_\mathrm{thres}$. The algorithm repeats itself from of point two, until the complete time series is processed. The bubble residence times per bubble event are now given by $T_{bubble}=T_{ET}-T_{AT}$, identical to the widths of the corresponding binary bubble mask, shown as the solid blue lines of figure \ref{fig:04}. Manual visual inspection of the algorithm shows a correct single-bubble-detection success rate of $>95\%$. The remaining 5\% are bubble events that perhaps should have been split in multiple separate bubble events, or vice versa.

The applied method as described above determines the latency length of our probe \emph{in situ} and by averaging over multiple bubble events that are deemed similar. The obtained value might differ from that obtained by experiments where a plane air-water interface is impacting the probe at predetermined velocities and attack angles, similar to the procedure as used by \cite{car92}. This would remove the influence of the (unknown) instantaneous bubble curvature and velocity as seen by the probe in the \emph{in situ} determination. The influence of these instantaneous differences might be reduced thanks to the averaging over multiple bubble events, but one cannot be sure unless tested. This is an interesting issue for future studies.

\subsubsection{Accuracy}
\label{sec:1pp_accuracy}
What can be said on the accuracy of the measurement technique? Obviously, there are many possible errors that are hard to quantify, but we can  capture the most prominent ones.

First of all, we must correct for the chance distribution of the frontal piercing location, leading to detected bubble chord lengths smaller than the major bubble chord. While this will not influence the detected local gas volume fraction, it will underestimate the bubble sizes. We quantify this underestimation in a first order approximation by assuming a spheroidal bubble with its frontal area as seen by the probe tracing a perfect circle of radius $R_1$ and with its third major chord parallel to the probe (i.e. in the azimuthal direction) of length $2R_2$. The mean detected chord length is then given by dividing the total volume of the spheroid $4/3 \pi R_1^2 R_2$ by its frontal area as seen by the probe $\pi R_1^2$, resulting in $4/3 R_2$, which is 1.5 times smaller than the major azimuthal chord of length $2 R_2$. Hence, we can expect a similar contribution to the underestimation of detected bubble sizes by a factor of $\xi_{chord} \approx 1.5$, as long as the bubbles can be sufficiently approximated by spheres, $R_1 = R_2$, or by spheroids, $R_1 \neq R_2$ (see figure \ref{fig:07} for photos of the studied bubbly flows). As will turn out in \S \ref{sec:bubble_size}, where we report on the mean aspect ratios of the bubbles as obtained from high-speed image analysis, the approximation would correspond to prolate spheroids, $R_2 > R_1$. We will take $\xi_{chord}$ into account when calculating and presenting bubble sizes. Studies on the (time-resolved) shape of individual bubbles fall outside of the scope of the current paper.

We can distinguish between two different error sources. One error source comes from the masking algorithm. Its ability to correctly detect single bubble events in time is key here. With only 5\% of wrongly segmented $T_{bubble}$ times, supposedly normally distributed around their correct values, the statistical mode of $T_{bubble}$ as used in our upcoming calculations, will change insignificantly when statistical convergence is achieved. Another factor concerning the algorithm is how the chosen threshold values are influencing the location of the retrieved time markers $T_{AT}$ and $T_{ET}$. This imposes an uncertainty error $\varepsilon_{mask}$ on the $T_{bubble}$ estimation. To our advantage here are the steep signal rates at the onset of tip dewetting (signal rise) and tip rewetting (signal fall). We estimate an error in time of $\pm 2$ samples, equivalent to $\varepsilon_{mask} = \pm 16.7$ $\mu$s. The quality of the bubble mask can be judged in figure \ref{fig:04}.

Another error source originates from the invasive nature of the optical fibre probe as studied by \cite{julia05}, possibly deflecting, decelerating and/or deforming the bubbles. As stated before, we are inclined to view the deflection and deceleration errors as negligible due to the large added mass of the bubbles. Nevertheless, we set the underestimation error in the local gas volume fraction due to bubble deflection to $\varepsilon_{deflect} = +5\%$. \cite{julia05} report an overall overestimation of $T_{bubble}$ caused by deceleration and an overall underestimation caused by the deformation, see figure 15 of \cite{julia05}. Hence, we set the error in $T_{bubble}$ due to deceleration at an overestimation of $\varepsilon_{decel.}=-5\%$. The error due to deformation we amply set at an underestimation of $\varepsilon_{deform}=+50\%$.

In the upcoming plots we will present error bars which are based on above error estimates. These error bars predominantly represent systematic errors. As proof for this statement, we refer the reader to appendix \ref{app:axial_depence} where we present excellent repeatability for the local measurements at $Re = 1.0 \times 10^6$.


\section{Global DR measurements} \label{sec:global_exp}


\subsection{Measurement procedure}

\begin{figure}
\center
\includegraphics[height=5.2cm]{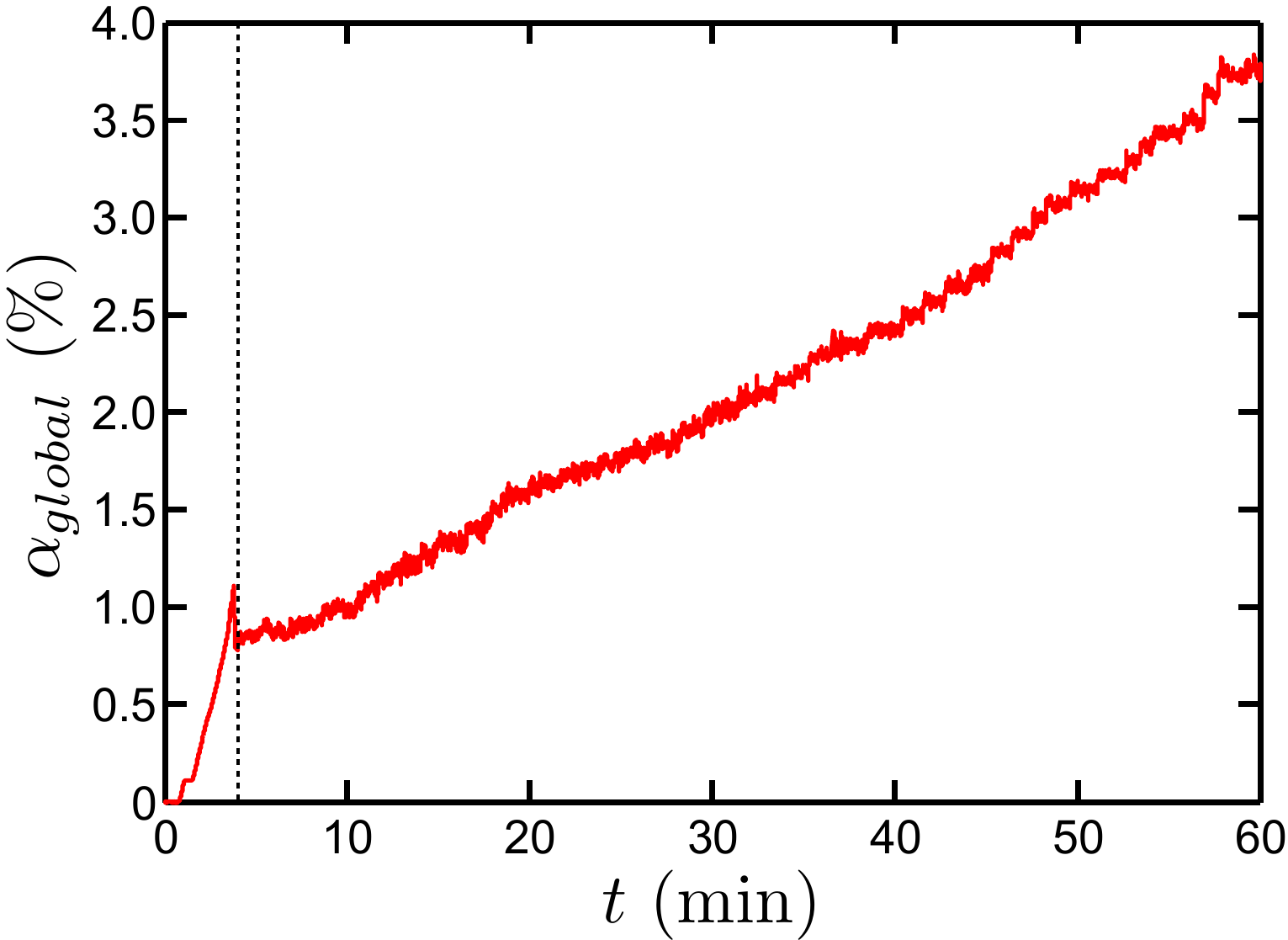}
\caption{Typical time evolution of the global gas concentration used in the global torque studies, shown here for the case of $Re = 5.1\times 10^5$. The sudden steep rise from $\gvf = 0\%$ to $1\%$ reflects the accumulation of bubbles into the flow below dynamic equilibrium, i.e.\ the flow is still able to draw in more bubbles
 than can escape. When the dynamic equilibrium is reached, here after approximately 4 minutes as indicated by the dotted vertical line, the gas injection rate is manually increased in a quasi-stationary way.
\label{fig:05}}
\end{figure}

We perform measurements on the global torque $\tau$ on the middle section of the IC as function of the global gas volume fraction $\gvf$ with the IC rotating at a fixed rate. We carry out experiments at $Re = 5.1 \x 10^5$, $1.0 \x 10^6$ and $2.0 \x 10^6$, corresponding to IC rotation frequencies of $\omega_i/2\pi = 5.0$, 10, and 20 Hz, respectively. We start each experiment with single phase flow and once temperature stability at $21 \pm 0.1^{\circ}$ C is achieved, the acquisition of the torque is initiated and the injection of air follows. We start with a small injection rate $\sim 1$ l/min. Initially, nearly all of the air bubbles become trapped in the strong turbulent fluctuations of the TC flow with little air escaping into the expansion vessel and hence the global gas volume fraction rapidly increases. Eventually, the flow becomes saturated with bubbles and a dynamic equilibrium settles in between the volume of air entering and leaving the flow, stabilizing $\gvf$. From now on we increase the injection rate gradually over time, forcing more air into the flow until we reach $\gvf \approx 4\%$. This ramp on $\gvf$ takes between 25 and 160 minutes, depending on the investigated $Re$ number and the gas injection rate. Higher-$Re$-number flow accumulates bubbles faster and hence $\gvf$ increases more rapidly. See figure \ref{fig:05} for a typical time evolution with the dotted vertical line representing the onset of dynamic equilibrium.


\subsection{Results}

\begin{figure}
\center
\includegraphics[height=7cm]{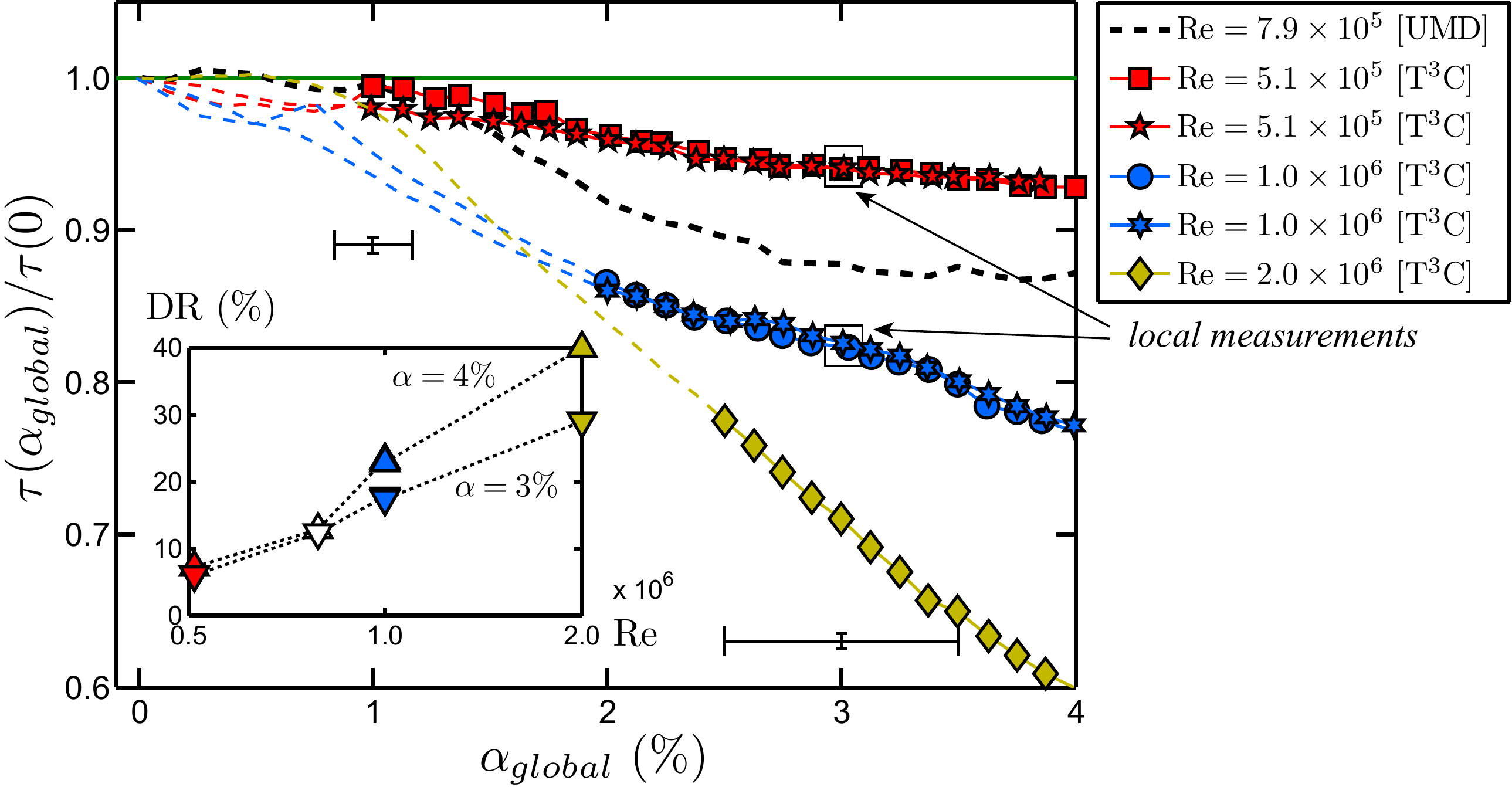}
\caption{Normalized dimensional torque $\tau$ as a function of $\gvf$. Closed symbols are data sets from the present experiments for various fixed $Re$ numbers. The dashed lines at the start of these sets indicate that the bubbly flow was not yet at dynamic equilibrium. The horizontal bars indicate the error of the gas concentration for the present measurements. Data are included from \citet{ber07} carried out on the University of Maryland (UMD) TC setup (fully dashed curve). Inset: drag reduction defined as the dimensional torque reduction ratio DR = 1 - $\tau(\gvf)/\tau(0)$ for various $Re$ numbers at constant $\gvf$ = 3\% (lower branch) and 4\% (upper branch). Colored symbols are the present results, and the open black symbols are results of \citet{ber07}.
\label{fig:06}}
\end{figure}

In figure \ref{fig:06} the dimensional torque data, normalized by the torque of the single phase flow $\tau(0)$, are plotted as a function of $\gvf$. Only after dynamic equilibrium has been reached in $\gvf$ do we represent the data by symbols instead of by dashed curves. The general trend of each experiment, carried out at a fixed rotation rate, shows a decrease in dimensional torque for increasing global gas volume fraction $\gvf$, when at dynamic equilibrium. Furthermore, the larger the $Re$ number, the more effective the reduction of the drag is. It is remarkable that there is a major DR of 40\% at $Re = 2.0 \x 10^6$ with $\gvf = 4 \pm 0.5\%$, which by far
was not attainable by previous measurements in turbulent TC flows.

In the figure unpublished data from \citet{ber07} have been included, carried out on the TC facility of the University of Maryland, or UMD TC for short. Their data, taken at $Re = 7.9 \x 10^5$, lie well between the present data of $Re = 5.1 \x 10^5$ and $1.0 \x 10^6$. Thus good agreement for bubbly DR has been found between the present measurements and those by \citet{ber07}. Also note the very good repeatability as demonstrated by the duplicated measurement series taken at $Re = 5.1 \x 10^5$ and $1.0 \x 10^6$.

One could argue that the reduction of torque can be explained by just a change of effective density and viscosity of the liquid due to bubble injection. In Appendix \ref{app:G_torque} we introduce a non-dimensional representation of the torque, trying to compensate for this density and viscosity dependence on the gas volume fraction. We demonstrate that this cannot account for the DR as observed in the strong DR regime by at least a factor of two in the $Re = 5.1 \x 10^5$ case and a factor of four in the $1.0 \x 10^6$ case. The bubbles must modify the flow actively in addition to simply changing the fluid properties.

A straightforward way to define the drag reduction is $\mathrm{DR} = 1 - \tau(\gvf)/\tau(0)$, where $\tau(\gvf)$ and $\tau(0)$ are the dimensional torque values of the two-phase and single-phase case, respectively. Using this DR definition, the net, practical effect on the torque is quantified. The inset of the figure shows this DR percentage for the data points at constant $\gvf = 3\%$ and 4\% as a function of $Re$. \Citet{ber05} find a pronounced DR for $Re \gtrsim 6 \x 10^5$ and a weak DR for $Re$ less than this. The present measurements show a similar result: strong DR for $Re = 1.0 \x 10^6$ and $2.0 \x 10^6$ and moderate DR for $Re = 5.1 \x 10^5$. We focus on the following two cases at $\gvf = 3 \pm 0.5\%$ for the local measurements: $Re = 5.1 \times 10^5$ with DR $\approx 6\%$ corresponding to the moderate DR regime and $Re = 1.0\times 10^6$ with DR $\approx 18\%$ corresponding to the strong DR regime.


\section{Local measurements} \label{sec:local_exp}
\begin{figure}
\center
\includegraphics[width=1\textwidth]{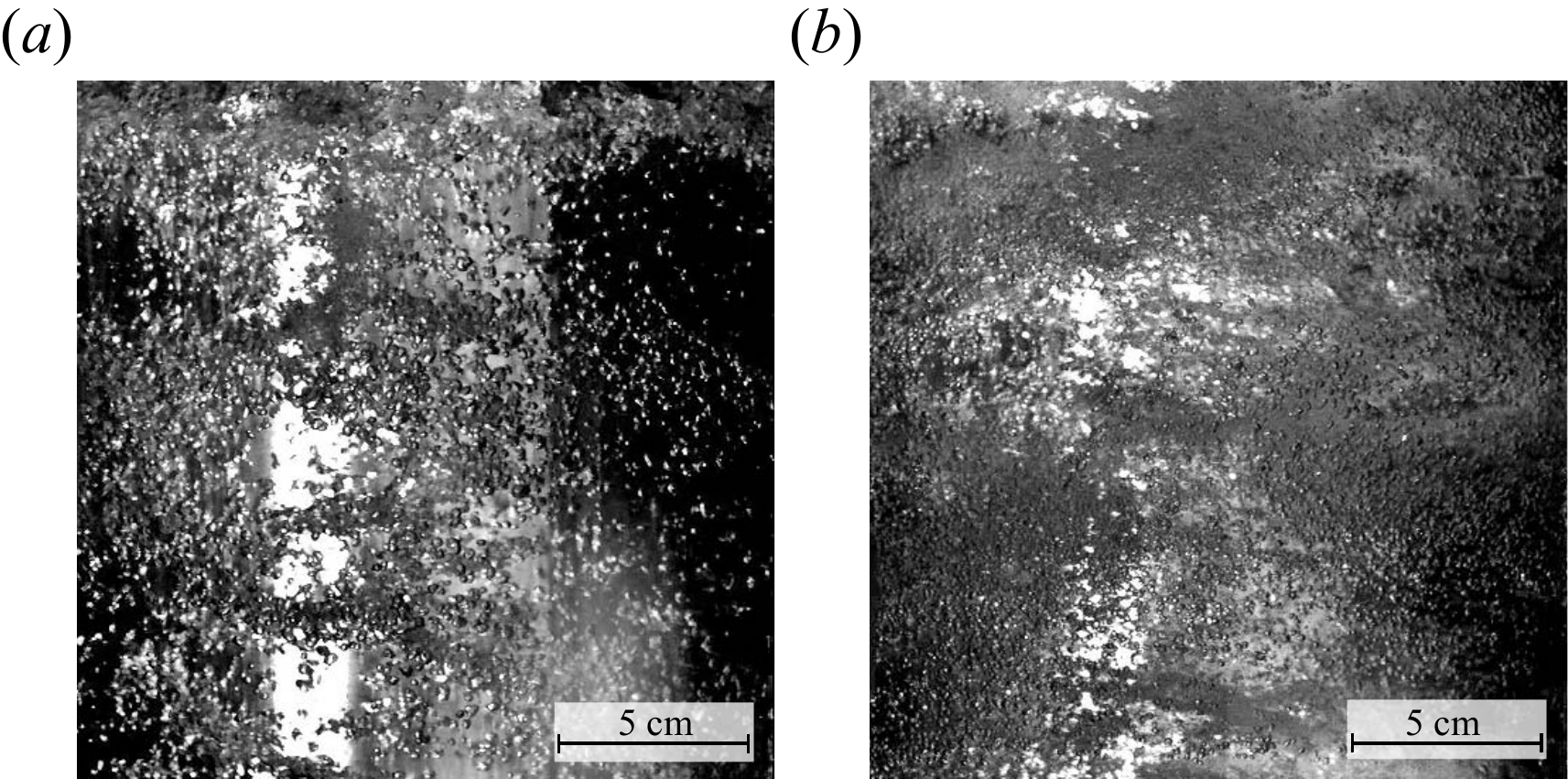}
\caption{High-speed images of the bubbly turbulent flow, taken through the transparent OC of the T$^3$C facility                 , for the two locally examined DR regimes at (\emph{a}) $Re = 5.1 \x 10^5$ and (\emph{b}) $Re = 1.0 \x 10^6$, both at $\gvf = 3 \pm 0.5\%$. Movies are available at \href{http://dx.doi.org/10.
1017/jfm.2013.96}{http://dx.doi.org/10.
1017/jfm.2013.96} included as supplementary material.
\label{fig:07}}
\end{figure}

We will study the local flow profiles and bubble statistics inside the TC gap for two different cases, selected out of the global DR results. They are $Re = 5.1 \x 10^5$, falling into the moderate DR regime, and $Re = 1.0 \x 10^6$, falling into the strong DR regime, both at a global gas volume fraction of $\gvf = 3 \pm 0.5\%$. To provide the reader with a qualitative sense of how such flows look, we present high-speed image snapshots of the two examined $Re$ cases in figure \ref{fig:07}.

We present local profiles on the azimuthal liquid velocity, estimated bubble size, gas concentration and Weber number, all scanned in the radial direction at middle height of the TC cell. Profiles at two other axial positions for the $Re = 1.0 \x 10^6$ case were also taken, which we present separately in Appendix \ref{app:axial_depence}.

\subsection{Measurement procedure}

The global gas volume fraction is maintained at a constant $\gvf = 3.0 \pm 0.5\%$ for both cases of constant $Re = 5.1 \x 10^5$ and $1.0 \x 10^6$, corresponding to IC rotation rates of 5.0 and 10 Hz, respectively. The temperature of the liquid is maintained at $21 \pm 0.5 ^{\circ}$C.

The local gas concentration profiles are scanned with the optical fibre probe for 300 s at each radial position. The measurement time of 300 s is chosen to ensure statistical convergence. The first radial position is as far from the IC wall as the traversing system allows, which is $\approx 27$ mm from the IC wall and roughly one-third into the gap towards the OC. In around 20 evenly radially distributed steps we scan up to the last radial position, which is as close as the probe's tip can safely be to the IC wall, i.e.\ down to 0.03 mm. The profile of $Re = 1.0 \x 10^6$ is measured twice to demonstrate repeatability.

The local azimuthal liquid velocity profiles are scanned by LDA in the single- and two-phase cases, and once more by particle image velocimetry (PIV) for the pure single-phase case (details may be found in \cite{hui12a}). Both techniques are set up to provide the full radial profiles, apart from the near-wall boundary layers. The profiles acquired with LDA consist of around 40 equally distributed radial positions. The number of detected Doppler bursts for each radial position ranges from $6 \x 10^3$ to $50 \x 10^3$ for $Re = 5.1 \x 10^5$ and from $28 \x 10^3$ to $50 \x 10^3$ for $Re = 1.0 \x 10^6$. The inner radial positions are increasingly harder to probe with LDA due to the visual shielding of the bubbles. Hence, we achieve an average data acquisition rate of $\approx 20$ Hz close to the IC, going up to $> 500$ Hz when entering the bulk flow region. The measurement time per radial position is checked to have enabled statistical convergence.


\subsection{Discussion of results}


\subsubsection{Azimuthal liquid velocity profiles}\label{sub:vel_profiles}

\begin{figure}
\center
\includegraphics[width=1\textwidth]{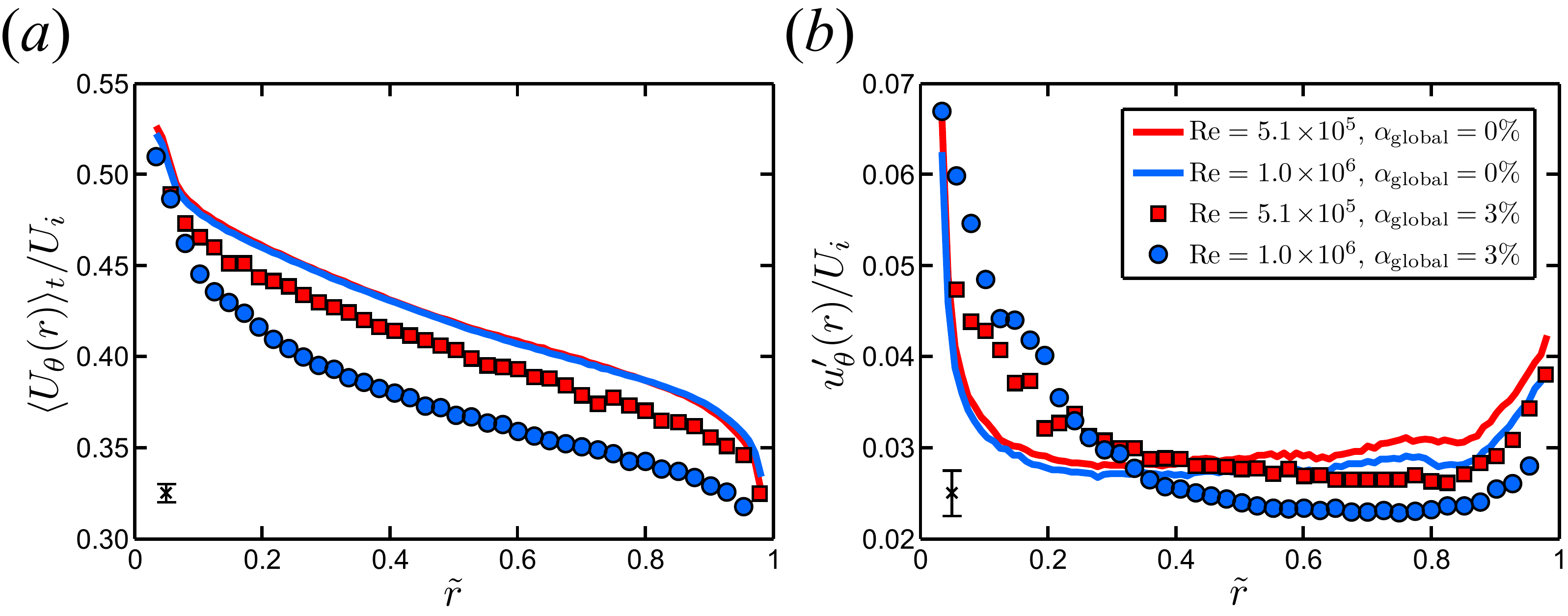}
\caption{The radial profiles of (\emph{a}) the azimuthal liquid mean velocity normalized by the velocity of the inner wall $U_i$ and (\emph{b}) the standard deviation of the normalized velocity fluctuations, both at middle height of the TC cell. The solid lines are obtained by PIV and the colored symbols are obtained by LDA.
\label{fig:08}}
\end{figure}

First, we study how the liquid velocity is modified by the injected bubbles in the two different $Re$ cases by comparing the single-phase case to the two-phase case. We measure the time-averaged local azimuthal velocity $\langle U_{\theta}(r) \rangle_t$ and the standard deviation of the local azimuthal velocity fluctuations $u'_{\theta}(r)$. The results are displayed in figure \ref{fig:08} with (\emph{a}) the mean velocity normalized by the IC wall velocity, which is $U_i = 6.28$ m/s for $Re = 5.1 \x 10^5$ and $U_i = 12.6$ m/s for $Re = 1.0 \x 10^6$, and (\emph{b}) the standard deviation of the normalized velocity fluctuations. The solid lines are the single-phase cases as measured by PIV and the colored symbols are data on the two-phase cases obtained by LDA. LDA measurements were also performed on the single-phase cases and they are found to be in good agreement with the PIV data: the mean within $\pm 0.6\%$ and the standard deviation within $\pm 5\%$ as indicated by the error bars in the lower left corner of the panels. To prevent the figure from cluttering up, we only show the PIV data for the single-phase cases. The single-phase LDA data of $Re = 1.0 \times 10^6$ can be found in appendix \ref{app:axial_depence}.

The rescaled mean velocity profiles in figure \ref{fig:08}(\emph{a}) fall on top of each other in the single-phase flow case for both $Re$ numbers. The bubbles clearly destroy this similarity. Since we do not have enough spatial resolution to resolve the boundary layers, we will focus on the bulk regime, taken as $0.2 < \tilde{r} < 0.8$. The mean azimuthal velocity profiles for both the single- and two-phase cases follow roughly the same shape in the bulk. The reduction of the bulk azimuthal velocity due to the bubble injection is $\approx 4\%$ at $Re = 5.1 \x 10^5$ and $\approx 12\%$ at $Re = 1.0 \x 10^6$. This is not unexpected as this velocity reduction should reflect the torque reduction ratio on the IC wall.

As shown in figure \ref{fig:08}(\emph{b}), the standard deviation of the normalized velocity fluctuations are nearly identical in the single-phase flow case for both $Re$ numbers. For the bubbly flow we observe an increase in $u'_{\theta}$ as compared with that of the single-phase flow at radial locations $\tilde{r} \lesssim 0.3$, i.e.\ close to the IC. For the higher-$Re$ case this increment is stronger. The next step is try to determine whether the local increase in the velocity fluctuations can be linked to the spatial distribution of the bubbles.


\subsubsection{Local bubble size} \label{sec:bubble_size}

\begin{figure}
\center
\includegraphics[width=1\textwidth]{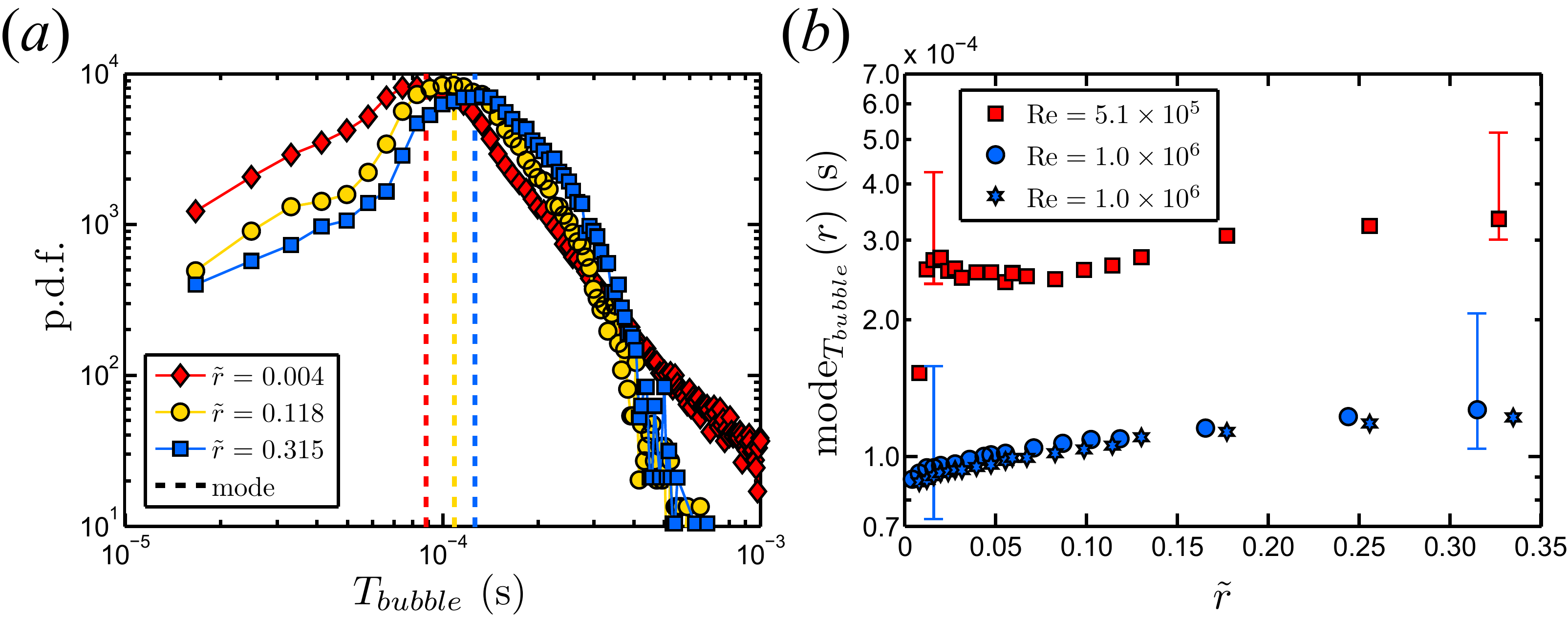}
\caption{(\emph{a}) Probability density functions of the bubble residence time for $Re = 1.0 \x 10^6$ at three non-dimensional gap distances. The mode of $T_{bubble}$ is indicated by the dashed vertical lines. (\emph{b}) The full radial profiles of the mode. The error bars show a strong underestimation, that is predominantly systematical as the repeatability for $Re = 1.0 \times 10^6$ is good.
\label{fig:09}}
\end{figure}

\begin{figure}
\center
\includegraphics[width=1\textwidth]{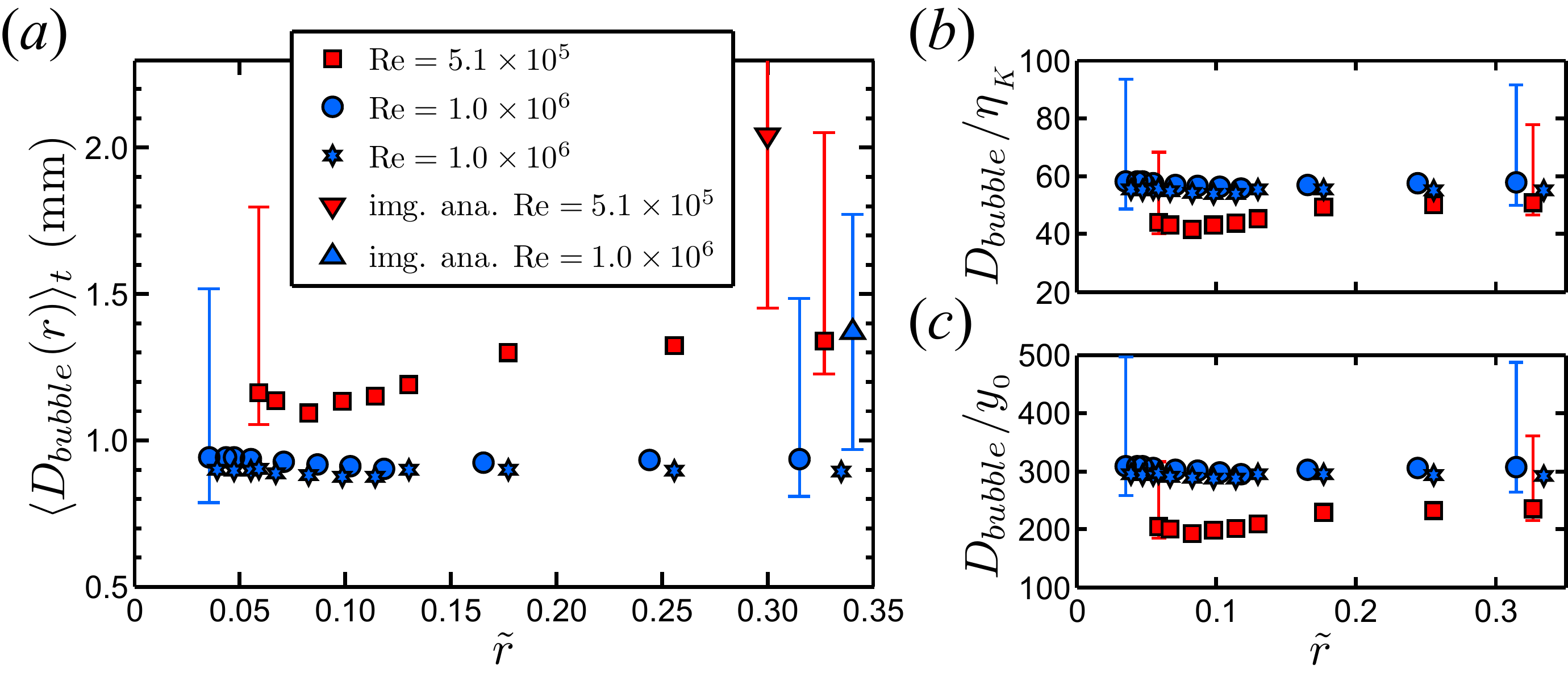}
\caption{(\emph{a}) Estimated mean horizontal bubble diameter obtained from the optical probe signals and the LDA velocity profiles. The triangles are the mean horizontal bubble diameters resulting from image analysis, with the error bars indicating the standard deviation. The size estimation from the optical probe is systematically underestimated by a factor of $\approx$ 1.5 for both $Re$ cases. (\emph{b}) The bubble diameter expressed in the Kolmogorov length scale $\eta_K$ and (\emph{c}) expressed in the wall unit $y_0$.
\label{fig:10}}
\end{figure}

Out of the fibre probe data we compute the histogram of the bubble residence time $T_{bubble}$ at each radial position. Figure \ref{fig:09}(\emph{a}) shows typical probability density functions (p.d.f.s) for three non-dimensional gap distances $\tilde{r}$ at $Re = 1.0 \x 10^6$. As can be appreciated, the mode of the p.d.f.s (defined here as the average of the bubble residence time corresponding to normalized probabilities higher than 0.85) is displaced the more to the right, the more distant the fibre tip is located from the IC. This is to be expected since the azimuthal flow velocity drops with increasing radius. The full radial profiles of the mode, $\mathrm{mode}_{T_{bubble}}(r)$, are displayed in figure \ref{fig:09}(\emph{b}). The error bars are based on the errors $\varepsilon_{mask}$, $\varepsilon_{decel.}$ and $\varepsilon_{deform}$ as discussed in \S \ref{sec:1pp_accuracy}. The furthest left data point for $Re = 5.1 \x 10^5$  in figure \ref{fig:09}(\emph{b}) is substantially below the rest of the profile, outside of the error bar range. The reason for this is yet unclear to us.

By multiplying, for each radial position, the mean azimuthal liquid velocity of the bubbly flow from figure \ref{fig:08}(\emph{a}) with the p.d.f.s mode of the bubble residence time, the mean horizontal bubble diameter as function of the gap distance is estimated: $\langle D_{bubble}(r)\rangle_{t} = \langle U_{\theta}(r) \rangle_t ~ \mathrm{mode}_{T_{bubble}}(r) ~ \xi_{chord}$. Several assumptions are applied here: (i) no slip or drift of the bubbles with respect to the surrounding liquid (to be examined in future work); (ii) the average of the instant $U_{\theta}T_{bubble} $ values is equivalent to the product of $\langle U_{\theta}\rangle_t$ with $\mathrm{mode}_{T_{bubble}}$; and (iii) the shape of the bubbles can be approximated by spheroids as described in \S \ref{sec:1pp_accuracy}, such that the detected chord lengths can be transformed to horizontal bubble diameters by applying the systematic $\xi_{chord}$ factor.

\begin{table}
\centering
\begin{tabular}[c]{c c c c c c c c}
$Re$ & $\varepsilon$ & $\eta_K$ & $\tau_w$ & $u_\tau$ & $y_0$\\
& (m$^2$ s$^{-3}$) & (m) & (N m$^{-2}$) & (m s$^{-1}$) & (m)\\
\\
$5.1\times10^5$ & ~1.96 & $2.63 \times 10^{-5}$ & ~29.7 & 0.172 & $5.69 \times 10^{-6}$ \\
$1.0\times10^6$ & 13.5~~ & $1.62 \times 10^{-5}$ & 102~~~ & 0.320 & $3.06 \times 10^{-6}$ \\
\\
\end{tabular}
\caption{Several flow parameters as calculated out of the globally measured torque for the two $Re$ numbers in the single-phase case: the energy dissipation rate $\varepsilon$, the Kolmogorov length scale $\eta_K$, the wall shear stress $\tau_w$, the friction velocity $u_\tau$ and the wall unit $y_0$.
\label{table:lengthscales}}
\end{table}

Figure \ref{fig:10}(\emph{a}) presents these estimated local mean bubble diameters, with the error bars based on $\varepsilon_{mask}$, $\varepsilon_{decel.}$ and $\varepsilon_{deform}$ originating from $T_{bubble}$. We report an overall larger mean bubble diameter of 1.2 (+0.7, -0.1) mm for the $Re = 5.1 \x 10^5$ case as compared with 0.9 (+0.6, -0.1) mm for the $Re =1.0 \x 10^6$ case due to the stronger shear in the latter. The size estimation of the bubbles is constant over the scanned radial distance for $Re = 1.0 \times 10^6$, exhibiting good repeatability (see the inset to figure \ref{fig:17}(\emph{b}) later for an exploded view of the data). This is a strong indication that the method has merit and that the errors are predominantly of systematic nature. For $Re = 5.1 \times 10^5$ the bubble size changes weakly over the radial TC gap. The reason for this is unknown, but it can be covered by a small random error.

We also use high-speed image recordings to measure the bubble size optically at one radial position, focusing at a distance $\tilde{r}\approx0.3$. The camera is pointed radially inwards, perpendicular to the curvature of the OC and the axial axis. For $Re = 5.1 \x 10^5$ we report a horizontal size of $2.0 \pm 0.6$ mm (mean $\pm$ standard deviation) and a vertical size of $1.6 \pm 0.5$ mm, taken over 234 in-focus bubbles. For $Re = 1.0 \x 10^6$ the horizontal size is $1.4 \pm 0.4$ mm and the vertical size is $1.1 \pm 0.3$ mm, taken over 144 in-focus bubbles. The triangles in figure \ref{fig:10}(\emph{a}) represent these horizontal bubble sizes with the vertical error bars as the standard deviation. It shows that the bubbles in the lower $Re$ case are less homogeneous in size as its standard deviation is larger than that for the higher $Re$ case. It also confirms the estimated horizontal bubble diameter from the optical fibre signals within the underestimation error bars. For both $Re$ cases the optical probe underestimates the horizontal bubble size at $\tilde{r}\approx0.3$ by a factor of $\approx$ 1.5. Finally, the mean aspect ratio, taken as the horizontal over the vertical bubble size, is 1.26 and 1.30 for $Re = 5.1 \times 10^5$ and $Re = 1.0 \times 10^6$, respectively, the latter exhibiting stronger overall bubble deformation. We use the term `overall' to stress that these aspect ratios are averaged over many different bubbles and, hence, do not reflect the time evolution of the shape of a single bubble.

We now compare in figures \ref{fig:10}(\emph{b}) and \ref{fig:10}(\emph{c}) the mean horizontal bubble diameter to the Kolmogorov length scale $\eta_K$ and the wall unit $y_0$, both calculated out of the globally measured torque of the single-phase flow at the corresponding $Re$ numbers. Table \ref{table:lengthscales} lists these values, together with other relevant flow parameters given for completeness, see \citet{lat92a,eck07b} for the definitions. In the radial region of $0.05 < \tilde{r} < 0.35$ the horizontal bubble diameter in the moderate DR case is around 46 (+25, -4) $\eta_K$ and 211 (+114, -19) $y_0$, and in the strong DR case it is around 56 (+34, -9) $\eta_K$ and 298 (+180, -46) $y_0$, in line to what has been reported by \citet{ber05}.


\subsubsection{Gas concentration profiles}\label{sec:gvf_profiles}

\begin{figure}
\center
\includegraphics[height=12cm]{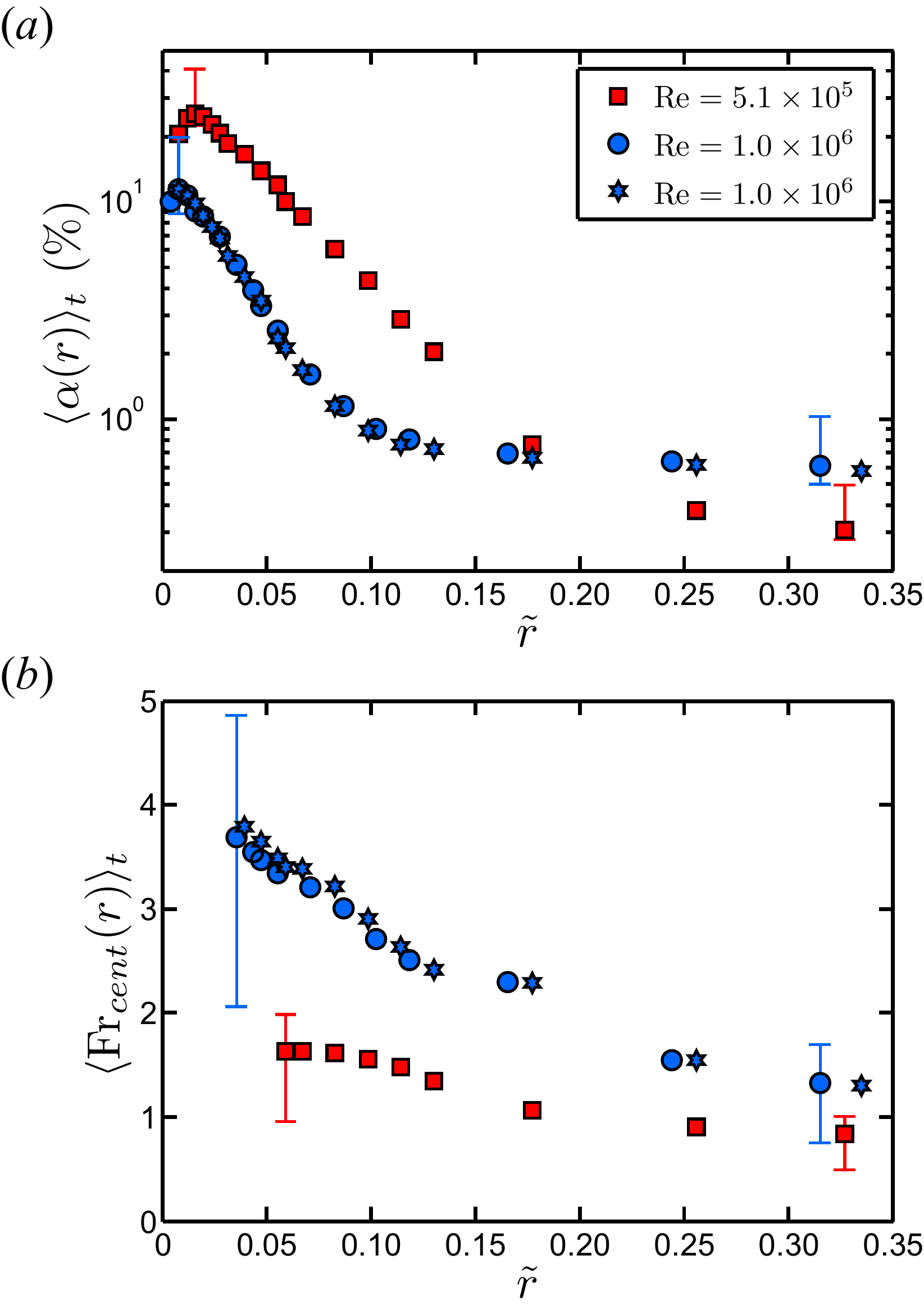}
\caption{Radial profiles at middle height of the two examined DR regimes: moderate DR at $Re = 5.1 \x 10^5$ and strong DR at $Re = 1.0 \x 10^6$, both at a global gas volume fraction $\gvf = 3 \pm 0.5\%$. (\emph{a}) Local gas concentration with a predominantly systemical underestimation. (\emph{b}) Local centripetal Froude number with a corresponding predominantly systematical overestimation. The repeatability for $Re = 1.0 \times 10^6$ is excellent.
\label{fig:11}}
\end{figure}

We calculate the time-averaged local gas volume fraction as a function of the radial position out of the bubble mask (see \S \ref{sec:fibre_probe_technique}) by summing over all bubble residence times and dividing this by the total measurement time, i.e.\
\begin{equation}
\langle \alpha(r)\rangle _t = \frac{\sum\limits_{i=1}^{N_{bubbles}} T_{bubble, i}}{T_{total}},
\end{equation}
with $N_{bubbles}$ the total number of detected bubbles. The resulting radial profiles on the gas concentration are shown as percentages in figure \ref{fig:11}(\emph{a}). The error bars are based on $\varepsilon_{mask}$, $\varepsilon_{decel.}$, $\varepsilon_{deform}$ and $\varepsilon_{deflect}$, resulting in a mainly systematical underestimation. Once again we stress the good repeatability as made apparent by the two independently measured and nonetheless collapsing profiles at $Re = 1.0 \x 10^6$.

Figure \ref{fig:11}(\emph{a}) clearly shows that the bubbles tend to accumulate close to the IC boundary regime for both $Re$ numbers. This phenomena is not surprising: in centrifugally driven flow there is, on average, a favorable pressure gradient pushing the bubbles towards the center of rotation, here with an acceleration $a_{cent}(r)=U_{\theta}^2/r$ equal to $\sim 3g$  for $Re = 5.1 \x 10^5$, and $\sim 9g$ for $Re = 1.0 \x 10^6$ (where $g$ denotes the gravitational acceleration). For this estimate we took the measured azimuthal liquid velocity at the center of the TC gap.

Next, we discuss the radial distance from the IC wall, $\ramax$, at which the gas concentration is maximum, $\amax \equiv \mathrm{max}\{\langle \alpha(r) \rangle_{t}\}$. For $Re = 5.1 \x 10^5$ this distance is $\ramax \approx 1.2$ mm, equivalent to $\tilde{r} \approx 0.016$. For the higher-$Re$ case it is located at $\ramax \approx 0.6$ mm, equivalent to $\tilde{r} \approx 0.008$. The value of $\ramax$ for each $Re$ number is close to the respective mean bubble diameter as presented in figure \ref{fig:10}(\emph{a}). However, from our measurements we cannot conclude that the $\ramax$ we measure is the correct radial distance at which the true maximum gas volume fraction occurs along the gap. It could be the case that the fibre probe, when located very near the IC wall, creates an up-flow high-pressure region that deflects the bubbles from their path, hence underestimating $\langle \alpha(r) \rangle _t$. In this region the corresponding $\varepsilon_{deflect}$ error might be of higher value than as set for the rest of the scanned radius. Further experiments are required to measure this aspect in detail.

More importantly, we compare the amplitude of the maximum local gas concentration for these two $Re$ numbers. For $Re = 1.0 \x 10^6$ we find $\amax \approx$ 11 (+11, -3) \%, whereas for $Re = 5.1 \x 10^5$ it arrives at about 25 (+22, -3) \%, which is 2.3 times higher. We evaluate the local Froude number to try to understand this difference.

If the flow were fully laminar then all of the bubbles would be pushed against the IC wall due to the radial pressure gradient associated with the centrifugally driven flow. With the addition of sufficiently strong turbulent liquid velocity fluctuations, the bubbles will experience pressure fluctuations, leading the bubbles to get spread also away from the wall into the bulk of the flow. We estimate the ratio of the pressure fluctuations' induced acceleration acting on the bubbles $a_{press~fluct}$ to the centripetal acceleration $a_{cent}(r)$ by introducing a centripetal Froude number,
\begin{equation}\label{eq:Fr_cent}
\Fr(r)=\frac{a_{press~fluct}}{a_{cent}(r)}=\frac{{u'}^2/D_{bubble}}{U_{\theta}^2/r},
\end{equation}
with $u'$ the standard deviation of the liquid velocity fluctuations, $U_{\theta}$ the mean azimuthal liquid velocity, $D_{bubble}$ the bubble diameter and $r$ the radial position to be considered.

For $u'$ and $U_\theta$ we take the measured azimuthal liquid velocity profiles of the two-phase LDA data of figure \ref{fig:08}, $u'_{\theta}(r)$ and $\langle U_\theta (r)\rangle_t$, respectively. \citet{burin10} report the same order of magnitude for the three cylindrical components of the liquid velocity fluctuations inside turbulent single-phase TC flow as measured by multi-component LDA. At this moment we can only assume that this similarity can also be applied to the two two-phase flow cases, but it should clearly be tested (the radial and axial components are hard to measure accurately as the flow field is dominated by the azimuthal component). For $D_{bubble}$ we take the time-averaged radial profile $\langle D_{bubble}(r)\rangle_t$ as shown in figure \ref{fig:10}. Because we use the local time-averaged profiles of each of these quantities, we retrieve the local time-averaged Froude number $\langle \Fr(r) \rangle _t$, presented in figure \ref{fig:11}(\emph{b}). The error bars are based on the small random errors in $\langle U_{\theta}(r)\rangle_t$ and $u'_{\theta}(r)$, together with $\varepsilon_{mask}$, $\varepsilon_{decel.}$ and $\varepsilon_{deform}$, arising from the $\langle D_{bubble}(r)\rangle_t$ estimation. The errors in the bubble size estimation are dominant and, hence, $\Fr$ is mainly systematically overestimated. For both $Re$ cases we find $\langle\Fr(r)\rangle_t \gtrsim$ 1, also within the error bars. Hence, the turbulent fluctuations are strong enough to diffuse the bubbles also against the centripetal force towards the bulk region of the flow, away from the inner wall. We do not claim that this evaluation is exact and it should be regarded as a first approach to try to explain the gas distribution differences.

Now we evaluate the local $\langle\Fr(r)\rangle_t$ number at the radial position $\ramax$ to understand the different $\amax$ for each $Re$ case. The closest we can get is $\tilde{r}= 0.06$ for which we find $\langle\Fr\rangle_t \approx$ 1.6 (+0.4, -0.7) for $Re = 5.1 \x 10^5$ and $\langle\Fr\rangle_t \approx$ 3.4 (+1.1, -1.5) for $Re = 1.0 \x 10^6$. The lower $Re$ case has a lower $\Fr$ number by a factor of $\approx$ 2.1, indicating that the effective centripetal force on the bubbles is higher and, hence, the bubble accumulation is stronger near the IC wall at the expense of a lower concentration in the bulk, as observed. The $\Fr$ number ratio at this radial position matches the maximum gas concentration ratio of $\approx$ 2.3 between both $Re$ cases.

The spatial broadening around the gas concentration maxima of figure \ref{fig:11}(\emph{a}) implies a wider bubble accumulation zone for $Re = 5.1 \times 10^5$ as opposed to $Re = 1.0 \times 10^6$. Clearly, there must be more at play than just the balance between pressure fluctuations and centrifugal forces on the bubbles for explaining the concentration profiles: if this would not be the case then the $\Fr$ number would dictate the narrowest spatial distribution to happen for $Re = 5.1 \times 10^5$.

What is unambiguously remarkable is the following: one would expect that the DR efficiency is correlated to the bubble concentration near the inner wall, and that a higher local gas volume fraction should give a larger DR. However, the present measurements show an opposite trend: the local gas concentration near the inner wall is higher in the moderate DR regime than that in strong DR regime. This clearly suggests that a higher local gas concentration near the inner wall does not guarantee a larger DR. What then is the dominant effect for the strong DR at the high $Re$ numbers? To try to answer this question we will look into bubble deformability by defining the local bubble Weber number.


\subsubsection{Local Weber number}

Following \citet{ber05}, we define the bubble Weber number as
\begin{equation} \label{eq:Weber}
\We = \frac{\rho {u'}^2 D_{bubble}}{\sigma},
\end{equation}
with $u'$ the standard deviation of the velocity fluctuations and $\sigma$ the surface tension of the bubble--liquid interface. Relation (\ref{eq:Weber}) is identical to what is used in \citet{ber05}. Some typographical errors appear in that work: the Weber number definition lists the radius of the bubble $R_b$ while it should read the bubble diameter $D_{bubble}$. This concerns their equation (4) and the Weber number definition three lines above it. Also, their abstract should read `$R$ = 0.5 mm' as the value of the used bubble radius and in the conclusion the statement `$R_b=42\eta =300y_0$' should read `$R_b=21\eta =150y_0$'. The rest of their values, including the right-hand side of equation (4) and their graphs are correct.

\begin{figure}
\center
\includegraphics[height=7cm]{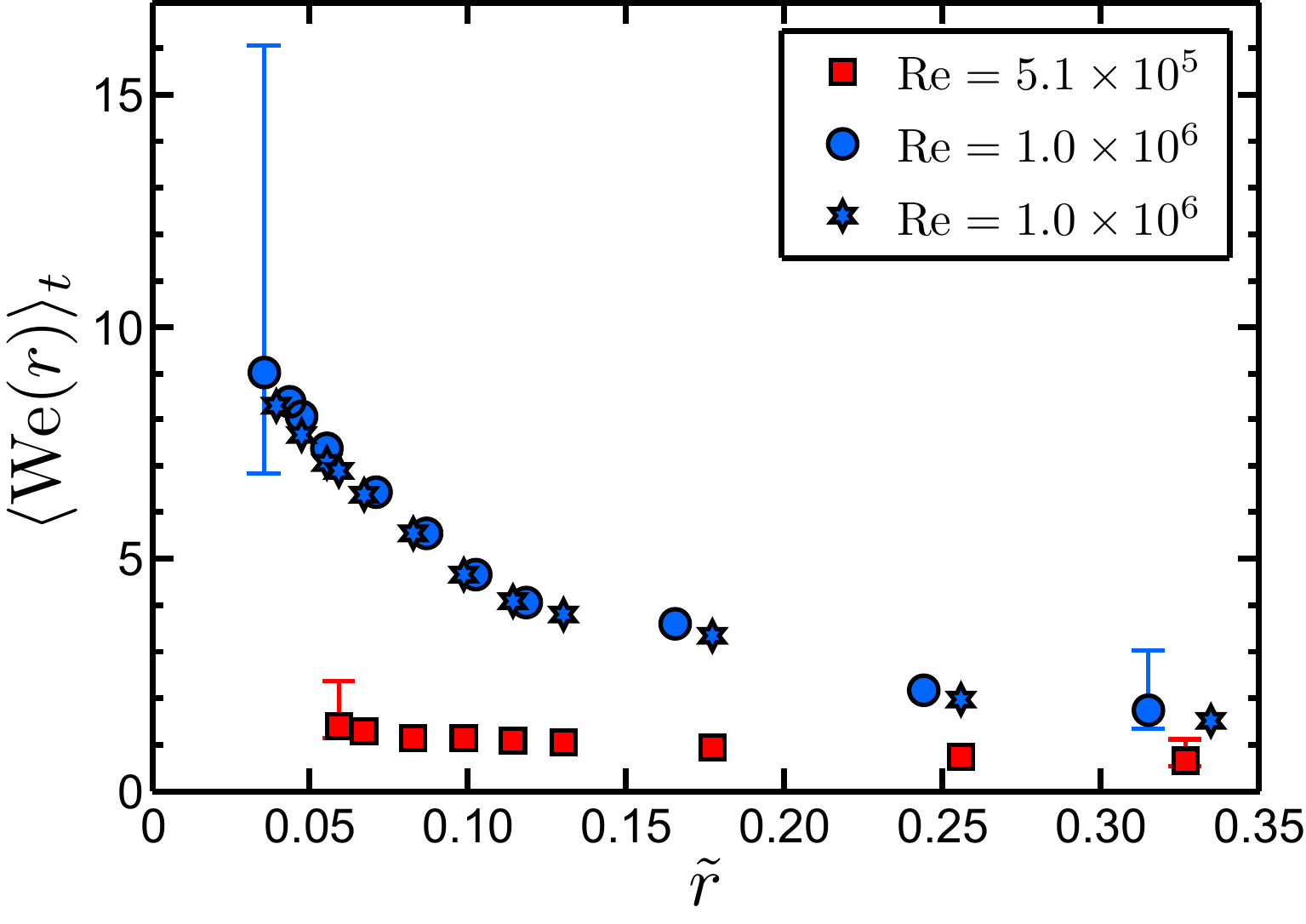}
\caption{Local time-averaged Weber number profiles of the two examined DR regimes at a global gas volume fraction of $\gvf = 3 \pm 0.5$\%: moderate DR at $Re = 5.1 \x 10^5$ and strong DR at $Re = 1.0 \x 10^6$. The vertical error bars indicate a predominantly systemical underestimation, originating from the bubble size estimation.
\label{fig:12}}
\end{figure}

For $u'$ we take the azimuthal liquid velocity fluctuations $u'_{\theta}(r)$ of the two-phase LDA data, see figure \ref{fig:08}(\emph{b}), following the same reasoning as used for the $\Fr$ number estimation in section \ref{sec:gvf_profiles}. For $D_{bubble}$ we use the local time-averaged bubble diameter $\langle D_{bubble}(r) \rangle_t$ of the optical fibre probe data, see figure \ref{fig:10}(\emph{a}). Substituting the values $\rho$ = 998 kg m$^{-3}$ and $\sigma$ = 0.0727 N m$^{-1}$, corresponding to water at 21$^{\circ}$C, provides us now with the local time-averaged Weber number of the bubbles inside the TC gap, scanned over one-third of the radius close to the IC at middle height of the TC cell. We present these $\langle \We(r) \rangle _t$ profiles for the two examined DR regimes in figure \ref{fig:12}. The vertical error bars are based on the small random error in $u'_\theta(r)$ together with $\varepsilon_{mask}$, $\varepsilon_{decel.}$ and $\varepsilon_{deform}$, arising from the $\langle D_{bubble}(r)\rangle_t$ estimation. The errors in the bubble size estimation are dominant and, hence, $\langle\We(r)\rangle_t$ is mainly systematically underestimated.

For $Re = 1.0 \x 10^6$ we see $\langle\We(r)\rangle_t > 1$ and that the bubbles increasingly deform when approaching the IC wall, where the liquid velocity fluctuations are stronger, finally leading to $\langle\We\rangle_t \approx 9$ (+7, -2) closest to the inner wall. The bubble Weber number for $Re = 5.1 \x 10^5$ is lower and it remains nearly constant with a value around 1, increasing to $\langle\We\rangle_t \approx 1.4$ (+1.0, -0.3) closest to the inner wall. This implies that the bubbles remain nearly spherical and/or retain their shape over time, because the velocity fluctuations are not strong enough to deform the bubbles, even though the estimated bubble size is larger, $\langle D_{bubble}(r)\rangle _t \approx$ 1.2 (+0.7, -0.1) mm, than that for the higher-$Re$ case, $\langle D_{bubble}(r)\rangle _t \approx$ 0.9 (+0.6, -0.1) mm. At $Re = 5.1 \x 10^5$ the global DR is around 6\%, whereas the DR is about 18\% for the case of $Re = 1.0 \x10^6$, having bubbles of higher deformability. Our local and global results thus clearly demonstrate that the bubble deformability is important for the observed strong DR, which is consistent with the findings by \citet{ber05} and \citet{ber07}, based on their global torque measurements, and by \citet*{lu05}, based on their numerical calculations. Indeed, it is this bubble deformation that is the origin of the lift force enhancement observed by \citet*{lu05}, resulting in the bubbles pushing away the liquid from the wall. Here we provide local experimental evidence on the important role of bubble deformability for strong DR in bubbly turbulent TC flow.


\section{Conclusion} \label{sec:conclusion}

We have shown that the net DR, based on global dimensional torque measurements in bubbly turbulent TC flow with IC rotation up to $Re = 2.0 \x 10^6$ and stationary OC, increases with increasing global gas volume fraction $\gvf \le 4\%$ and that the efficiency of the DR increases for increasing $Re$ numbers. We find a remarkable net DR of 40\% obtained at $\gvf = 4 \pm 0.5\%$ for $Re = 2.0 \x 10^6$, which was not attainable by previous TC set-ups. The present global measurements at $Re = 5.1 \x 10^5$ and $1.0 \x 10^6$ show good agreement with previous global measurements from \citet{ber07}. We observe a moderate DR for $Re = 5.1 \x 10^5$ up to 7\%, and a strong DR for $Re = 1.0 \x 10^6$ and $2.0 \x 10^6$ up to 23\% and 40\%, respectively.

To understand the flow differences between the moderate and strong DR regimes, we studied the flow locally for two $Re$ numbers at fixed $\gvf = 3 \pm 0.5\%$, each falling into a different DR regime. The smaller one, $Re = 5.1 \x 10^5$, represents the moderate DR regime at DR = 6\% and the larger one, $Re = 1.0 \x 10^6$, represents the strong DR regime at DR = 18\%. Consequently, we have extended the global flow experiments by \citet{ber05, ber07} by providing local flow information.

We investigated the azimuthal liquid velocity in the single- and two-phase flows for each $Re$ case. The amplitude of the mean azimuthal velocity in the bulk ($0.2 < \tilde{r} < 0.8$) for the bubbly flow decreases $\approx 4\%$ at $Re = 5.1 \x 10^5$, and $\approx 12\%$ at $Re = 1.0 \x 10^6$ as compared with that of the single-phase case. These reduction ratios match the global torque reduction ratios. Furthermore, we measured increased liquid velocity fluctuations close to the IC in both two-phase cases as compared with their equivalent single-phase case.

For retrieving local information on the bubble statistics we made use of the optical fibre probe technique. We have discussed the possible error sources of this technique and we have demonstrated that it underestimates the bubble size, predominantly due to systematic errors, by a factor of $\approx 1.5$ in our examined $Re$ cases as compared to high-speed image analysis.

The local gas concentration measurements show that the bubbles tend to accumulate close to the IC boundary regime for both $Re$ numbers. This can be explained by the centrifugally driven flow, pushing the bubbles towards the IC wall with an acceleration of around 3 to 9 $g$. The velocity fluctuations are strong enough to diffuse the bubbles also towards the bulk of the flow, as expressed by the centripetal Froude number $\Fr \gtrsim 1$. The radial distance from the IC wall at which we found the maximum local gas concentration is close to the respective bubble diameter. Surprisingly, the maximum local gas concentration at $Re = 5.1 \x 10^5$, i.e.\ in the moderate DR regime, is $\approx 2.3$ times higher than that at $Re = 1.0 \x 10^6$, i.e.\ in the strong DR regime. We found agreement between the time-averaged local $\langle\Fr(r)\rangle_t$ number and the maximum gas concentration near the inner wall, explaining this large concentration difference in a first approach. More importantly: this result unambiguously reveals that a higher local gas concentration near the inner wall does not guarantee a larger DR.

By estimating the time-averaged local Weber number $\langle\We(r)\rangle_t$ of the bubbles, we concluded that in the strong DR regime at $Re = 1.0 \x 10^6$ the larger liquid velocity fluctuations deform the smaller bubbles of $\langle D_{bubble}(r)\rangle _t \approx$ 0.9 (+0.6, -0.1) mm, since $\langle\We(r)\rangle_t > 1$. The local Weber number increases with decreasing distance from the inner wall, reaching $\langle\We\rangle_t \approx$ 9 (+7, -2) close to the inner wall, implying that the deformation of the bubbles is stronger near the IC wall. In contrast, in the moderate DR regime at $Re = 5.1 \x 10^5$ the bubbles remain close to spherical and/or retain their shape over time more, because $\langle\We(r)\rangle_t \sim 1$ with near the inner wall $\langle\We\rangle_t \approx$ 1.4 (+1.0, -0.3), even though the bubbles are larger in size: $\langle D_{bubble}(r)\rangle _t \approx$ 1.2 (+0.7, -0.1) mm.

We conclude that increasing the gas volume fraction has a positive influence on the azimuthal liquid velocity decrease, which is responsible for the observed DR. What is important for strong DR is that bubbles deform ($\We \gtrsim 1$) close to the boundary layer regime of the IC wall, leading to a large azimuthal velocity decrease and thus a strong DR. These local results match and extend the conclusion on the global torque results as found by \citet{ber05} and \citet{ber07}, and on the numerical results by \citet*{lu05}, both stressing the central role of bubble deformability.

\section{Discussion \& outlook}\label{sec:discussion}

\begin{figure}
\center
\includegraphics[width=.8\textwidth]{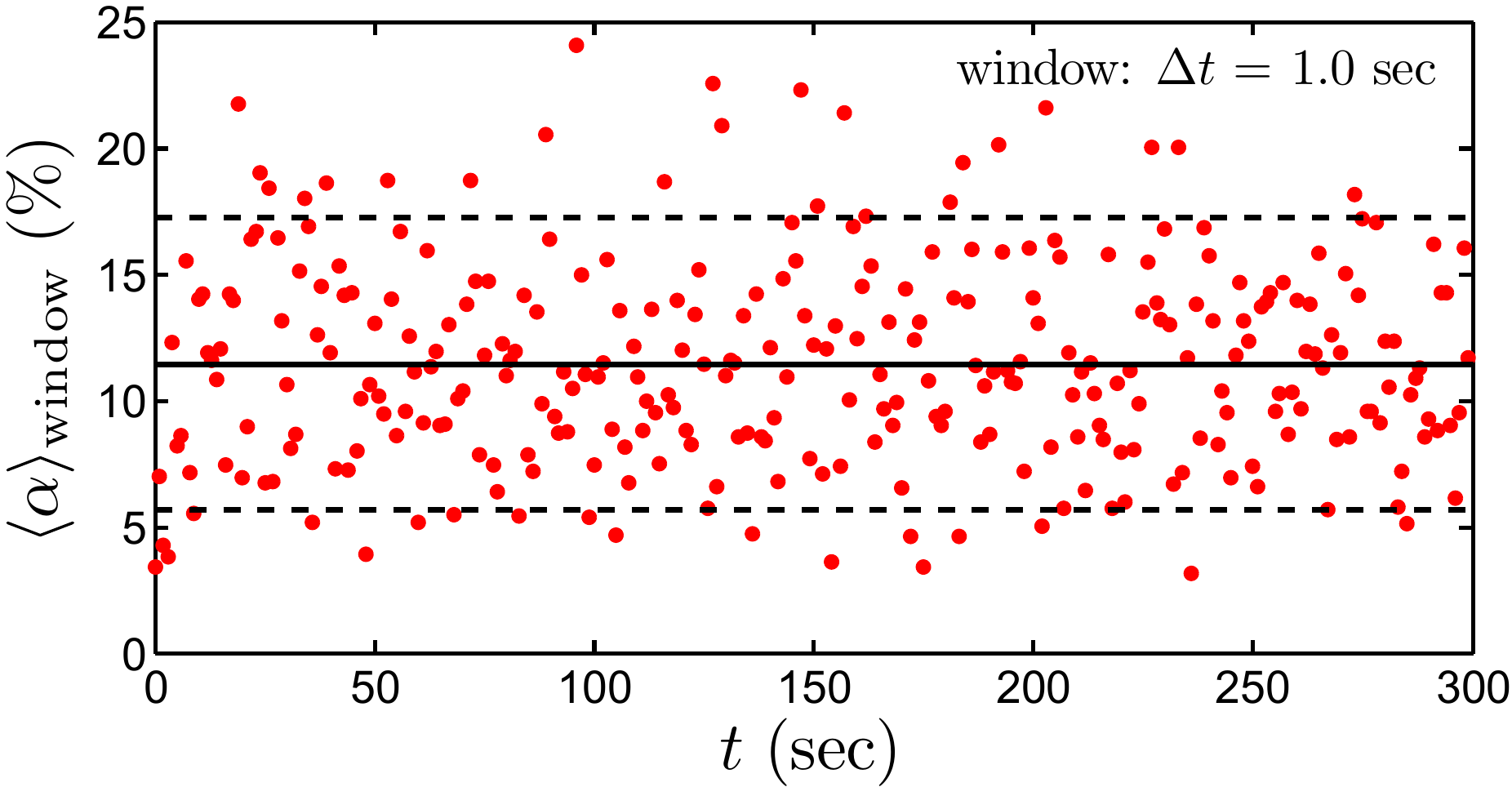}
\caption{Typical time series of the measured local gas volume fraction, here post-processed with a sliding average window of 1 s in duration, showing the short-time fluctuations and long-time stability. The circles are the windowed (quasi-instantaneous) gas volume fraction, the solid vertical line is the mean and the dashed vertical lines are the standard deviation of the fluctuations around the mean. The time series correspond to the $Re = 1.0 \times 10^6$ measurement (stars) at the radial location of maximal gas volume fraction $\ramax$ of figure \ref{fig:11}(\emph{a}).
\label{fig:13}}
\end{figure}

No periodicity has been found in the bubble signal time series at any of the examined radial locations. This includes spectral and auto-covariance analysis on bubble and inter-bubble arrival times. For completeness, we show in figure \ref{fig:13} a typical time series of the measured local gas volume fraction, here post-processed with a sliding average window of 1 second in duration, revealing the short-time fluctuations and long-time stability.

For the $Re = 5.1 \times 10^5$ case we occasionally saw bubbles coalesce into air patches of several centimeters in size. Sometimes, these air patches traveled along the inner wall, supporting the notion of a lubricating gas film. Mostly, however, the air patches seemed to be inside of the bulk, albeit with strong upward motion. For $Re = 1.0 \times 10^6$ the observation of air patches is even less frequent, and if they exist then they are more often observed to be inside of the bulk and not in contact to the wall. These events correspond to the long right tails of the bubble residence time p.d.f.s and their probability is 3 orders of magnitude smaller than that of the mode. Hence, we are of the opinion that the occasional air patches do not influence our results noticeably.

We note that the bubble distribution is a two-way coupled problem: bubbles are drawn into a radial distribution as suggested by $\Fr(r)$ and at the same time the magnitude of $\Fr(r)$ is determined by the velocity fluctuations induced by the presence of the bubbles. Likewise, the effect of bubble deformability also influences the velocity fluctuations' strength. The effect of deformability can be eliminated by injecting rigid neutrally buoyant particles into the flow, similar to what was used in experiments by \citet{ber05}. One could also further study above problem by changing the $\Fr$ and $\We$ numbers independently of each other. This can be achieved by changing the liquid density $\rho$ as it enters the $\We$ definition but not the $\Fr$ definition. Note that changing the surface tension $\sigma$ will not result in independently changing $\Fr$ from $\We$ as a different surface tension results in a different mean bubble diameter, which enters both the $\Fr$ and $\We$ definitions again.

In this present work the bubble diameter is linked to the driving control parameter $Re$, because the shear strength of the flow dominates the bubble size \citep{ris98}. One can achieve a different scaling of the $Re$ number to the bubble size by adding a low concentration of surfactant into the bubble carrier liquid. Even small concentrations of surfactant are known to significantly increase the effective surface tension of the bubble--liquid interface as it results in changing the bubble surface boundary condition from a slip to a no-slip condition, see e.g.\ the review article by \cite{tak11}. Different low concentrations of different surfactants can result in a variety of mean bubble sizes and deformability, and hence different $\We$. Of course, the huge shear strength of our flows will still play a role. Surfactants also prevent bubble coalescence and lead to a more monodisperse bubble size distribution, both aiding the simplicity of the flow. We assume that adding small amounts of surfactants will not change the turbulent flow properties, which should be tested experimentally. The bubble deformability and the connection to the global torque DR can and should be studied further in bubbly turbulent TC flow.

So far we have examined bubbly turbulent TC flow by only rotating the IC and keeping the OC at rest. The flow profile inside the TC gap becomes completely different once OC rotation is activated, see \cite{gil12} for angular velocity profiles of such flows obtained in the same TC facility. In the co-rotation regime the flow will become less turbulent and is hence not the first choice to examine. In the case of counter-rotation the flow will experience more shear and a `neutral line', i.e.\ the radial position where the mean angular velocity is zero, will exist inside the TC gap, thereby changing the overall radial pressure gradient induced by the centrifugal force. This could have a pronounced effect on the bubble distribution, possibly allowing indirect control of the radial position of maximum gas concentration $\ramax$ from the boundary layer region towards the bulk region.

Another route to proceed is to work towards a one-to-one comparison between our experiments and the front-tracking numerical simulations by Tryggvason and coworkers for the TC geometry, which, at least for lower $Re$ numbers, should be achievable.


\section*{Acknowledgments}

This study was financially supported by the Technology Foundation STW of the Netherlands. The authors would like to thank A. Chouippe for several high-speed imaging measurements and S.G. Huisman for the PIV measurements; R. Delfos, M.J.W. Harleman, C. van der Nat, G. Ooms, T.J.C. van Terwisga, E. van Rietbergen and J. Westerweel for scientific discussions; R.F. Mudde and J. van Raamt for help on the optical probe. We also gratefully acknowledge
continuous technical support by B. Benschop, S.J. Boorsma, M. Bos, G.W.H. Bruggert and G. Mentink.


\appendix


\section{Non-dimensional torque reduction ratio} \label{app:G_torque}

\begin{figure}
\center
\includegraphics[height=7.3cm]{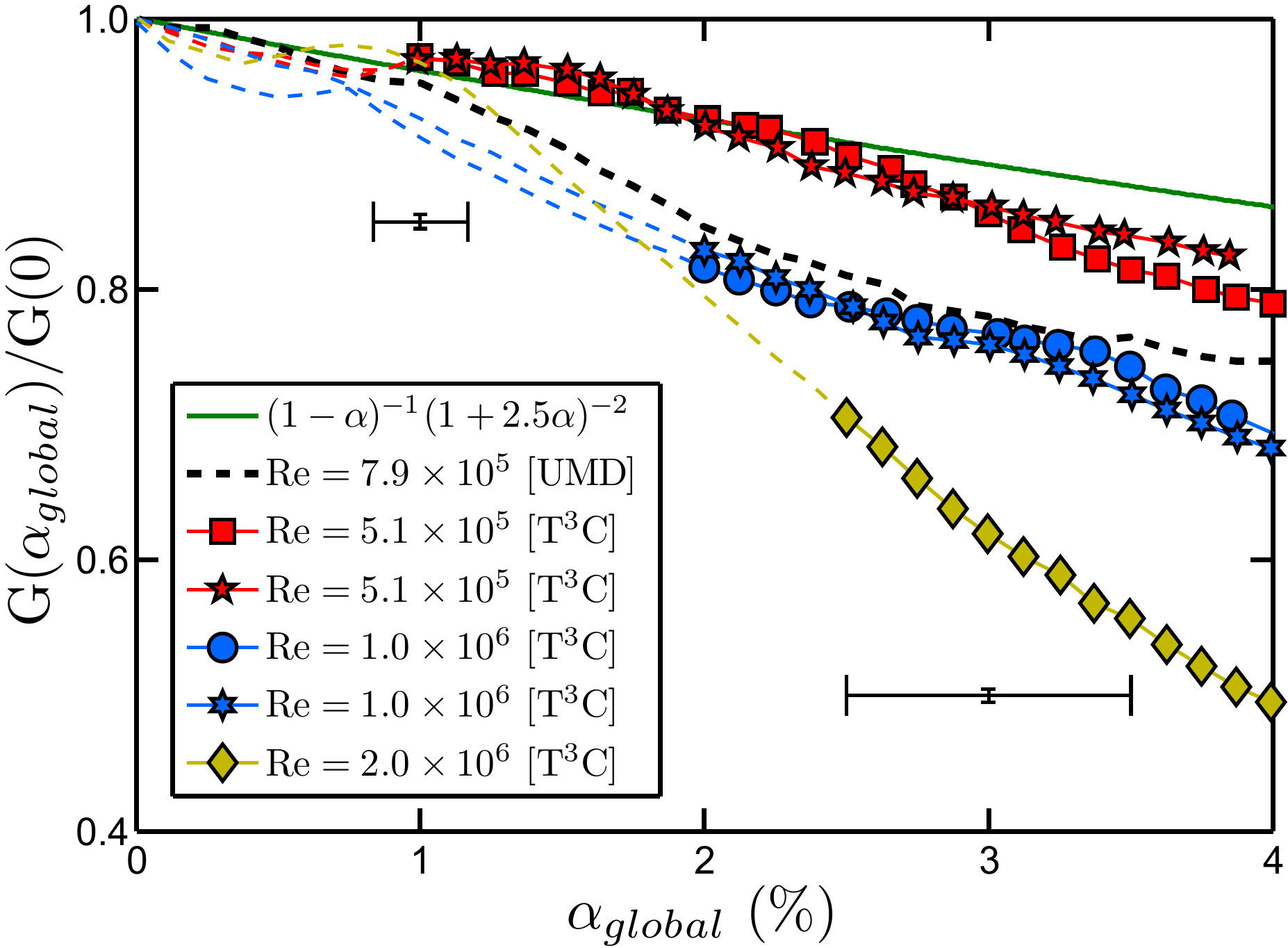}
\caption{Normalized non-dimensional torque $G$ as function of $\gvf$. This figure shows an alternative representation of the data in figure \ref{fig:06}. The solid green curve shows the suggested dependence of the density and viscosity terms of the normalized non-dimensional torque on $\gvf$.
\label{fig:14}}
\end{figure}

When injecting bubbles into a liquid at small concentrations, the effective kinematic viscosity and density of the two-phase fluid can be approximated, respectively, as follows:

\begin{eqnarray}
\rho_{\gvf} = \rho (1 - \gvf),\\
\nu_{\gvf} = \nu (1+ \frac{5}{2}\gvf), \label{eq:einstein}
\end{eqnarray}

\noindent where $\gvf$ is the global gas volume fraction, and $\rho$ and $\nu$ are the density and viscosity of the liquid at $\gvf = 0$. The `Einstein' equation (\ref{eq:einstein}) \citep{ein1906} is a first order approximation only and assumes the dispersed phase to be non-self-interacting and to be of the same density as the carrier phase. Although the applicability becomes questionable for bubbly turbulent flows with up to 4\% gas concentration, we will use (\ref{eq:einstein}).

To estimate the magnitude of the torque reduction, purely caused by the change of fluid properties when injecting bubbles into the flow, we define a non-dimensional torque based on the two-phase viscosity and density as
\begin{equation}
G(\gvf) =\frac{\tau}{2 \pi \rho_{\gvf} \nu_{\gvf}^2 L_{mid}}  =\frac{\tau}{2 \pi \rho(1-\gvf) \nu^2 (1+ \frac{5}{2}\gvf)^2 L_{mid}}, \label{eq:G}
\end{equation}
with $\tau$ as the measured torque on the middle section of the IC. We focus on the effect of bubbles on the torque and neglect the change of the $Re$ number.

The experimental results on the non-dimensional torque reduction are plotted in figure \ref{fig:14}, comparable to the dimensional torque presented in figure \ref{fig:06}. The reason for the shift in the UMD data can be contributed to a different fluid temperature of 24 $^{\circ}$C. The solid green curve represents the suggested dependence of the density and viscosity terms of the normalized non-dimensional torque on $\gvf$. Therefore, the difference between the solid green curve and each of the measured curves tells us the contribution to the DR that is not solely caused by a change of fluid properties. In the case of $Re = 5.1 \x 10^5$ the difference remains very small even up to $\gvf\approx 2.5\%$, but it increases significantly when $\gvf > 2\%$ for $Re = 1.0 \x 10^6$ and $2.0 \x 10^6$. Hence, the bubbles in the strong DR regime must modify the flow actively in addition to simply changing the fluid properties.


\section{Axial dependence at $Re = 1.0 \x 10^6$} \label{app:axial_depence}

\begin{figure}
\center
\includegraphics[width=.6\textwidth]{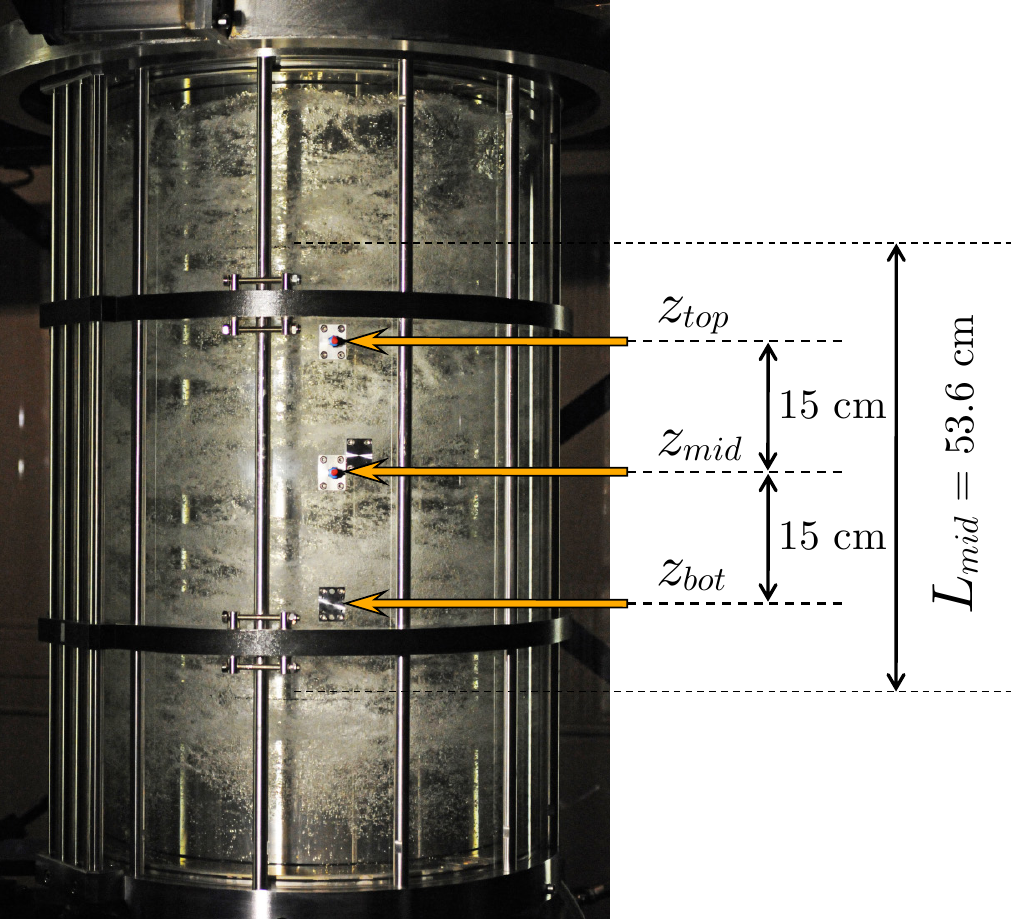}
\caption{Snapshot of the T$^3$C setup at $Re = 1.0 \x 10^6$ and $\gvf$ = 3\%. The axial measurement locations are tagged. Possible remnants of turbulent Taylor vortices are visible, that seem to coincide with the measurement locations. This is only coincidental at the moment of the snapshot. Neither stable vortices nor statistically stationary vortices have been observed in the bubbly flow by means of high-speed image analysis.
\label{fig:15}}
\end{figure}

We also investigate the axial dependence of the flow at $Re = 1.0 \x 10^6$ with $\gvf = 3 \pm 0.5\%$. The OC provides three locations through which the optical probe can be inserted. These are located at $z_{mid}=L/2$, $z_{bot}=z_{mid}-15$ cm and $z_{top}=z_{mid}+15$ cm, see figure \ref{fig:15}. In this appendix we will omit showing the error bars on the bubble statistics as they are of predominantly systematical nature and, hence, do not aid the axial dependence investigation.

\begin{figure}
\center
\includegraphics[width=1\textwidth]{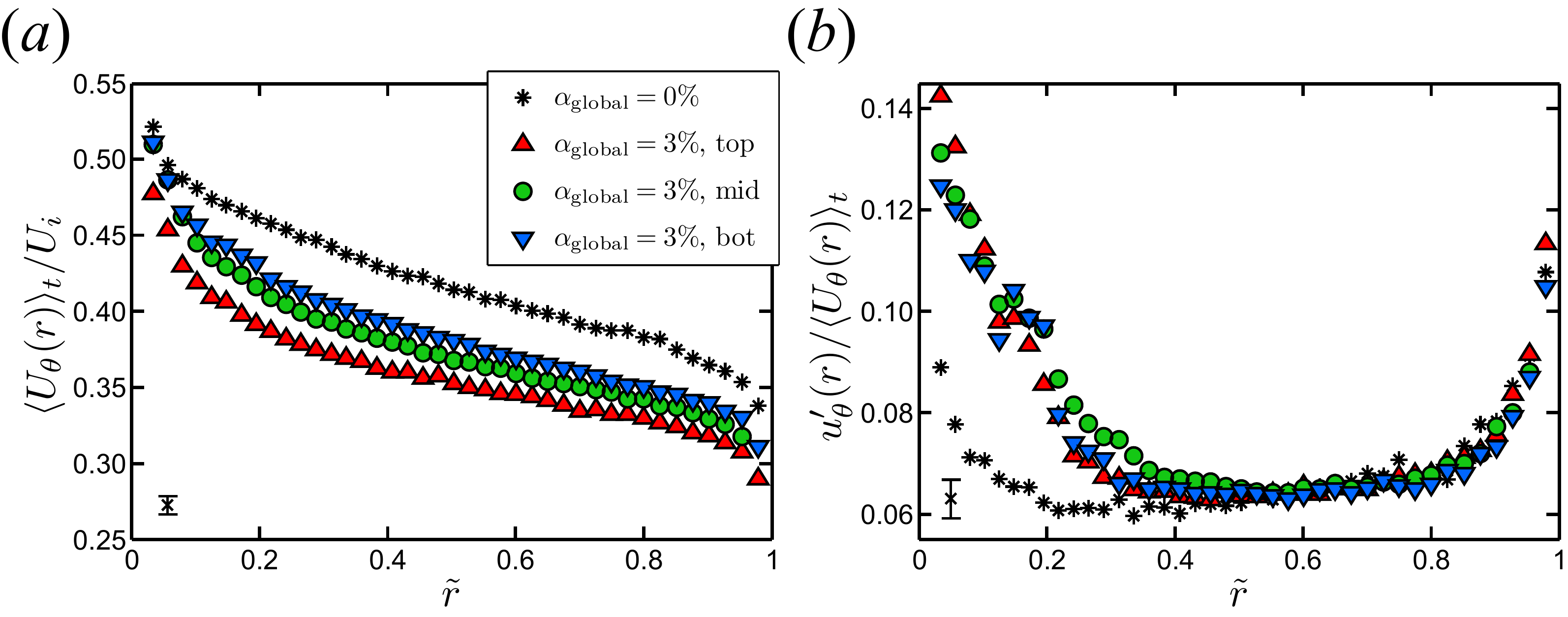}
\caption{The radial profiles as measured by LDA for $Re = 1.0 \times 10^6$ at different axial locations of (\emph{a}) the liquid azimuthal mean velocity normalized by the IC wall velocity $U_i$ and (\emph{b}) the turbulence intensity. The error bars in the lower-left corners indicate the errors on the data series.
\label{fig:16}}
\end{figure}

We will first focus on the azimuthal liquid velocity profiles as measured by LDA at these locations: the normalized mean velocity $\langle U_{\theta}(r)\rangle _t/U_i$ in figure \ref{fig:16}(\emph{a}) and the turbulence intensity $u'_\theta(r)/\langle U_\theta(r)\rangle_t$ in figure \ref{fig:16}(\emph{b}). For the single-phase case we find no axial dependence over the examined range as both the mean velocity and turbulence intensity profiles are identical to within 0.7\%. Hence we represent the single-phase case as the black stars. When we inject bubbles into the flow we observe a weak axial dependence on the mean azimuthal velocity but not on the turbulence intensity, see the colored symbols. The relative mean velocity amplitude difference in the bulk of the flow for the $z_{top}$ and $z_{bot}$ profiles is $\pm 4\%$ when compared with the $z_{mid}$ profile. Can this be linked to the bubble distribution?

\begin{figure}
\center
\includegraphics[width=1\textwidth]{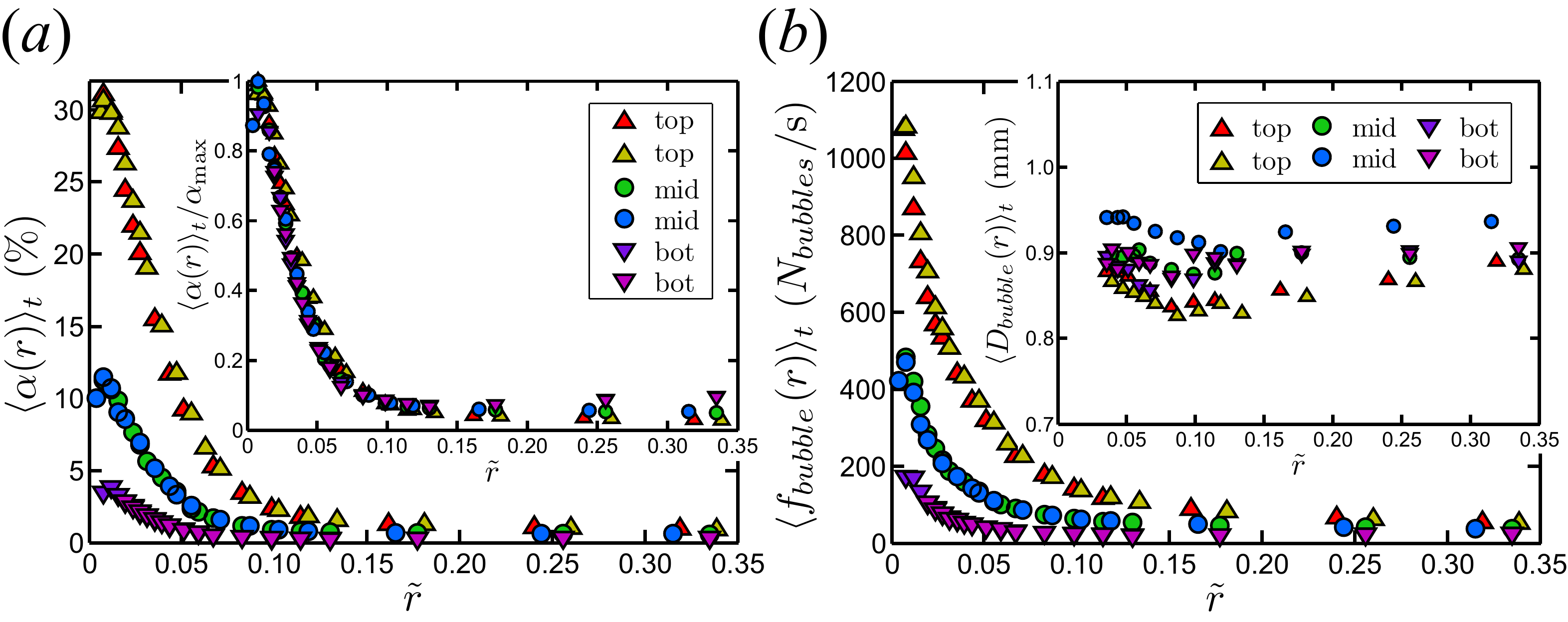}
\caption{(\emph{a}) Local gas concentration profiles at three axial locations in the flow at $Re = 1.0 \x 10^6$ and $\gvf$ = 3\%. Inset: the local gas concentration profiles normalized by their respective maxima $\amax$, hence revealing universality. (\emph{b}) The local average bubble detection rate in number of bubbles $N_{bubbles}$ per second. Inset: corresponding local mean bubble diameter. \label{fig:17}}
\end{figure}

So, we examine the corresponding gas concentration profiles, see figure \ref{fig:17}(\emph{a}). Two independent measurements are performed at each height to show the measurement repeatability. We note an increase in the maximal value of the local gas concentration $\langle \alpha(r) \rangle_t$ for increasing height. Between $z_{top}$ and $z_{mid}$ and between $z_{mid}$ and $z_{bot}$ there is a factor $\approx 2.8$ difference. These increments in gas volume fraction can be due to two factors: more bubbles or bigger bubbles.

Hence, we calculate, in the same manner as described in \S \ref{sec:bubble_size}, the local bubble diameter individually for each axial position, see the inset of figure \ref{fig:17}(\emph{b}). No axial dependence on the bubble size is found. When we consecutively calculate the average bubble detection rate $\langle f_{bubble}(r)\rangle_t$ for each axial position, we indeed find a match to the maximum gas volume fractions, see figure \ref{fig:17}(\emph{b}). Interestingly, the local bubble distribution profiles show universality when normalized by their respective maxima $\amax$, see the inset of figure \ref{fig:17}(\emph{a}).

The reason for this strong axial gradient on the gas volume fraction becomes apparent when exploring the following force balance. We find that the bubbles experience a centripetal acceleration $U_{\theta}^2/r \approx 9g$ in the bulk of the flow as compared to the $1g$ of axial upward acceleration due to buoyancy. These factors are not well enough apart to neglect the influence of gravity on the gas distribution. This can also explain why possible remnants of turbulent Taylor vortices, as might be seen in figure \ref{fig:15}, are not able to maintain a stable axial position, because they are being disrupted by rising bubbles. The contribution of these fluctuating large scale vortical structures to the measured standard deviation of the liquid velocity fluctuations remains unknown and should be examined in future work.

We conclude that the local gas concentration depends strongly on the axial position. However, the liquid mean velocity profiles change only weakly and the turbulence intensity profiles fall on top of each other. This suggests that the azimuthal turbulence intensity is not sensitive to the local gas concentration along the axial direction.


\bibliographystyle{jfm}

\end{document}